\documentclass[11pt]{amsart}
\usepackage{amsbsy,amssymb,amscd,amsfonts,latexsym,amstext,delarray, amsmath,graphicx,color,caption}
\usepackage{tikz}
\usetikzlibrary{matrix,arrows}
\usepackage{graphicx}
\input xy

\xyoption{all}
\usepackage{euscript}
\usepackage{multirow}
\usepackage{etex, pictexwd,dcpic}

\def\to{\longrightarrow}

\newcommand{\Z}{\mathbb{Z}}

\newtheorem{theorem}{Theorem}[section]
\newtheorem{proposition}[theorem]{Proposition}
\newtheorem{lemma}[theorem]{Lemma}

\newtheorem{corollary}[theorem]{Corollary}

\newcommand{\eqnlab}[1]{\label{eqn:#1}}

\def\g{\mathfrak{g}}

\def\C{{\mathbb C}}

\def\R{{\mathbb R}}
\def\Z{{\mathbb Z}}

\def\g{{\mathfrak g}}

\def\a{\alpha}

\def\Mg{\mathfrak{g}}

\begin{document}

\title{\bf {Integral forms of Kac--Moody groups and \\ Eisenstein series in low dimensional \\ supergravity theories}}

\date{\today}

\author{Ling Bao and  Lisa Carbone}

\vspace{5.0cm}

\begin{abstract} Kac--Moody groups $G$ over $\mathbb{R}$ have been conjectured to occur as symmetry groups of supergravities in dimensions less than 3, and their integer forms $G(\mathbb{Z})$ are conjecturally U--duality groups. Mathematical descriptions of $G(\mathbb{Z})$, due to Tits, are  not amenable to computation or applications. We describe a construction of  Kac--Moody groups over $\mathbb{R}$ and $\mathbb{Z}$ by Carbone and Garland, using an analog of Chevalley's constructions in finite dimensions and Garland's constructions in the affine case. We review a construction of Eisenstein series on finite dimensional semisimple algebraic groups using representation theory, which appeared in the context of superstring theory, and indicate how to extend it to general Kac--Moody groups. This coincides with a generalization of Garland's Eisenstein series on affine Kac--Moody groups to general Kac--Moody groups and includes Eisenstein series on $E_{10}$ and $E_{11}$. For finite dimensional groups, Eisenstein series encode the quantum corrections in string theory and supergravity theories. Their Kac--Moody analogs will likely also play an important part in string theory, though their roles are not yet understood.

\end{abstract}

\thanks{The first author and second authors were supported in part by NSF grant number DMS--1101282.}

\maketitle

\newpage
\tableofcontents
\section{Introduction}

Let $\mathfrak{g}$ be a symmetrizable Kac--Moody Lie algebra  of {\it {finite, affine, or hyperbolic} } type, over  a field $K$, and let $G$ be a Kac--Moody group associated to $\mathfrak{g}$. If $\mathfrak{g}$ is of finite type, then $\mathfrak{g}$ is a finite dimensional semisimple Lie algebra, and $G$ is a semisimple Lie group. Many of these occur as symmetries in physical models such as gravity, supergravity and supersymmetric gauge theories.

 Kac--Moody groups and algebras are the most natural extensions to infinite dimensions of finite dimensional simple Lie groups and Lie algebras. Affine Kac--Moody algebras and their generalizations by Borcherds have concrete physical realizations and have wide applications in physical theories. 

 Suitable extensions of the Dynkin diagrams of affine Kac--Moody algebras give rise to hyperbolic and Lorentzian Kac--Moody algebras. Recently  hyperbolic and Lorentzian Kac--Moody groups and algebras have been discovered as symmetries in certain physical models, though their mathematical properties and their role in theoretical physics is not fully understood. 
 
 There is evidence however, to suggest that groups and algebras of type $E_{10}$ and $E_{11}$ appear as symmetries of eleven dimensional supergravity ([DHN1], [W1]). Moreover, non--holomorphic automorphic forms on these groups are of interest in extensions of the Langland's program  \cite{BFH,Sh}. They are conjectured  to encode  higher derivative corrections of string theory and M--theory ([DN2], [DHHKN], [W1]) and automorphic forms on $E_{10}(\Z)$ have been conjectured to play a role in string theory ([BGH], [Ga]). These groups and automorphic forms are the objects of our study.

 It is known that the maximal supergravity theory in $D=11-n$ dimensions, which is the low energy effective action of type II string theories in $D$ dimensions, exhibits $E_{11-n}(\mathbb{R})$ symmetry ([CJ]). This theory admits charged solitonic solutions which are quantized by a type of Dirac quantization condition.  The presence of solitons  indicates that  $E_{11-n}$  symmetry occurs over $\mathbb{Z}$.

 We thus seek a description of the integral form, or $\mathbb{Z}$-form, of a Kac--Moody group $G$ over $\mathbb{R}$, analogous to $SL_2(\mathbb{Z})$ in $SL_2(\mathbb{R})$. In the affine case, Garland gave a construction of $G(\Z)$ ([G1]).

In the non--affine case,  Tits' group functor on the category of commutative rings ([Ti]) should be suitable for our purposes, but Tits' definition, by generators and relations, is very complicated. In the case of $E_{10}$, the problem of finding the full set of defining relations is intractable ([A3]).

 A further difficulty is that there is no unique definition of a Kac--Moody group, but rather several constructions using a variety of techniques and additional external data. For example, Kac and Peterson constructed Kac--Moody groups using integrable  modules ([KP]), but their construction works only over $\mathbb{C}$ and does not give rise to $G(\mathbb{Z})$. Moreover, their construction uses all integrable modules simultaneously, which makes applications technically cumbersome.

 In this review paper, our objective is to develop a more concrete approach to Kac--Moody group theory, with a view towards physical applications and particular emphasis on Kac--Moody groups over $\mathbb{Z}$. We describe a construction of [CG2] of Kac--Moody groups over $\R$ and $\Z$ using an analog of Chevalley's constructions in finite dimensions.  Our approach follows Garland's construction in the affine case ([G1]), with the necessary generalizations to the non--affine case.

The classical Eisenstein series on the upper half plane converges (in a half space) to an eigenfunction of the Laplacian that is invariant under the modular group $SL_2(\mathbb{Z})$ and its congruence subgroups.  In greater generality, automorphic forms may be constructed on $K\backslash G(\R)/G(\Z)$, for $G$ a semisimple algebraic group and $K$ a maximal compact  subgroup of $G$. They appear naturally in the context of superstring theories, in particular, they encode the quantum corrections in the low energy regime described by the supergravity theories ([GMRV], [LW]). The seminal examples are the maximal supergravity theories, which can be obtained by toroidally compactifying eleven dimensional supergravity. The global symmetry groups $G$ that appear in this case are the noncompact real forms of exceptional Lie groups $E_n(\R)$. Here, by $E_n(\R)$ we mean the noncompact split real form of $E_n$, often denoted $E_{n(n)}$.

 The scalar fields of the maximal supergravity theory in $(11-n)$ dimensions take values in the coset $K(E_{n(n)}) \backslash E_{n(n)}$. These coset symmetries have been shown to hold for maximal supergravity theories in ten dimensions all the way down to two dimensions via dimensional reduction ([CJ], [Ju1], [N]).

 By continuing dimensional reduction to one dimension, there is indication that the Kac--Moody group $E_{10}$ is  a symmetry group of eleven dimensional supergravity ([Ju2]). The role played by $E_{10}$ in eleven dimensional supergravity has been further investigated  by studying supergravity near a spacelike cosmological singularity ([DHN1]). A correspondence between the fields of the supergravity theory and the $E_{10}$ coset model was established after certain truncations were made on both sides of the correspondence ([DHN2]). In particular, this correspondence holds only at `low levels' of the roots of the Kac--Moody algebra.

 Lorentzian $E_{11}$ Kac--Moody symmetry appears in the study of eleven dimensional supergravity. In particular, West showed in [W2] that truncated versions of the bosonic sectors of eleven dimensional supergravity and type IIA supergravity can be derived as non--linear realizations in terms of certain truncation of $K(E_{11}) \backslash E_{11}$.

 Although the symmetry of the classical maximal supergravities involves Lie groups constructed over the real numbers $\R$, adding quantum corrections breaks these continuous groups to certain discrete subgroups called  U--duality groups ([HT]). For example, the subgroup  $E_{10}(\mathbb{Z})$ of the hyperbolic Kac--Moody group  $E_{10}(\mathbb{R})$ is conjectured to be a U--duality
symmetry group of Type II superstring theory ([Ju1]).  In analogy with $SL_2(\mathbb{Z})$, the group $E_{10}(\mathbb{Z})$ is also conjectured to be a `modular group' for certain automorphic forms that are expected to arise in the context of eleven dimensional supergravity ([DN]).

The physical scattering amplitudes of a string theory can be grouped in an elegant manner using automorphic forms defined on the double coset $K(G(\R))\backslash G(\R)/G(\Z)$, where $K$ is the maximal compact subgroup of $G$. In general, the automorphic forms that appear in string theory  transform as representations of $K$. However, due to the difficulty in analyzing them, most of the work so far has been conducted on  Eisenstein series which are invariant under $K$. 
 
 We note  that if $G$ is a Kac--Moody group, the  U--duality group $G(\Z)$ is not discrete in the usual sense, and the `unitary form' $K$ is not compact, though it is the analog of a maximal compact group in the finite dimensional case.


 Eisenstein series on the finite dimensional Lie groups that appear in the maximal supergravities have been studied systematically. Garland extended the classical theory of Eisenstein series to `arithmetic  quotients'  $K\backslash G(\R)/G(\Z)$\footnote{To prove convergence,  Garland extended $G$ by the automorphism $e^{-rD}$, where $r>0$ and $D$ is the degree operator.}
  of affine Kac--Moody groups $G$. We extend this work, with suitable modifications, to non--affine Kac--Moody groups such as $E_{10}$ and $E_{11}$.

 Braverman and Gaitsgory developed a theory of `geometric Eisenstein   series'  for affine Kac--Moody groups in the framework of the geometric Langlands correspondence ([BG]). This involves a geometric reformulation of number theoretic and representation theoretic notions in terms of algebraic curves and vector bundles.

 We define Eisenstein series on $K\backslash G(\R)/G(\Z)$, for $G$ a non--affine Kac--Moody group. Roughly speaking, the method is to appeal to Iwasawa decomposition $G(\R)=KA^+N$, construct a discrete eigenfunction (quasi--character) on the subgroup $A^+$ and then extend it to the whole of $G(\R)$ via Iwasawa decomposition, which is given uniquely. We then average over an appropriate coset space to obtain a $G(\Z)$-invariant function on $K\backslash G(\R)/G(\Z)$, where $K$ is the fixed point subgroup of the involution on  $G(\mathbb{R})$ induced from the Cartan involution on the Kac--Moody algebra $\mathfrak{g}(\mathbb{R})$. As we show, this is analogous to the construction of Eisenstein series on $SL_2(\R)$. 

 To make this method more transparent, we invoke highest weight modules for Kac--Moody algebras. This amounts to an extension of the construction of Eisenstein series on simple algebraic groups using representation theory, to Kac--Moody groups.

 In some cases, certain constraints are required on Eisenstein series defined using representation theory in order that these automorphic forms are eigenfunctions of the Laplacian. This will be discussed briefly throughout, though many questions remain open, even in the finite dimensional case (see for example [OP1]).

 For our applications, we will also require a fundamental domain for $G(\mathbb{Z})$ analogous to the fundamental domain for $SL_2(\mathbb{Z})$ on the Poincar\'e upper half plane $SL_2(\mathbb{R})/SO_2(\mathbb{R})$. We outline the results of [CGP] in this direction.

 The authors would like to thank  Pierre Cartier, Thibault Damour,  Steve Miller, Manish Patnaik, Boris Pioline and Pierre Vanhove for useful and enlightening discussions. We are indebted to Howard Garland for checking the details of our work, for his patient and helpful explanations and for introducing us to the subject. We are grateful to Peter West for clarifying the details of his results, for several improvements to the manuscript and for many helpful discussions. The early part of this work was done while the first two authors were at IH\'ES. We take pleasure in thanking the IH\'ES for its hospitality and wonderful work environment. The first author would also like to thank the Issac Newton Institute for its hospitality in Spring 2012 where part of this work was completed.

\newpage 
\section{Groups over $\Z$}

\subsection{Chevalley basis and Chevalley's construction of $SL_2$}\label{SL2Chev}

 The discrete group $SL_2(\Z)$ can be described as the group of 2$\times$2 matrices of determinant 1 with $\Z$-entries.  For exceptional Lie groups and infinite dimensional Kac--Moody groups,  in the absence of a suitable matrix representation, another method is required to define these groups over $\Z$. 

 As a motivating example, we describe Chevalley's construction of $SL_2$ in such a way that the group can be defined over $\C$, over $\R$ and over $\Z$. A benefit of this approach is that it gives us a set of generators and relations for these  groups. 

 This method can be generalized to Kac--Moody groups as we shall see. We shall use the following external data associated with the Lie algebra $\g=\mathfrak{sl}_2$:

  (1) A $\Z$-form $\mathcal{U}_{\mathbb{Z}}(\g)$ of the universal enveloping algebra $\mathcal{U}(\g)$.

  (2) A $\Z$-form $V_{\Z}$ of a highest weight module $V$ for $\g$.

   To construct $\mathcal{U}_{\mathbb{Z}}(\g)$ for $\g=\mathfrak{sl}_2$, we make use of a {\it Chevalley basis}.  

 Let $\g=\g_{\C}$ be any finite dimensional semisimple Lie algebra over $\C$. Let $\mathfrak{h}$ be a Cartan subalgebra and let $\Delta$ be the roots of $\g$ relative to $\mathfrak{h}$. Let $\ell=dim_{\C}(\mathfrak{h})$, $I=\{1,\dots,\ell\}$ and let $\Pi=\{\a_1,\dots ,\a_{\ell}\}$ be the simple roots.

 A {\it Chevalley basis} for  $\Mg$ is a basis
$$\mathcal{B}=\{x_{\a},h_i\mid \a\in\Delta,\ x_{\a}\in\mathfrak{g}_{\alpha},\ h_i\in\mathfrak{h},\ i=1,\dots ,\ell\},$$
for $\mathfrak{g}$ with
$$[h_i,h_j]=0$$
$$[h_i,x_{\a}]=\langle\a,\a_i\rangle x_{\a}$$
$$[x_{\a},x_{-\a}]={\alpha}^{\vee}\in \mathfrak{h},$$
and such that there is a  linear map  
$$\theta:\Mg\to\Mg$$ 
which preserves  $\mathfrak{h}$ and which takes  $x_{\a}$ to $x_{-\a}$ for all $\a\in\Delta$.

The existence of a Chevalley basis uses the following.

\begin{lemma} ([Bou]) Let $\Mg$ be a semisimple Lie algebra. There is a unique automorphism $\theta\in Aut(\Mg)$  which is equal to $-1$ on $\mathfrak{h}$ and which takes $\Mg_{\a}$ to $\Mg_{-\a}$ for all $\a\in\Delta$.
\end{lemma}

The automorphism $\theta$ is called the {\it Chevalley involution}.

 The usual basis 
 $e=x_{\alpha}= \left(\begin{matrix} 0 & 1\\ 0 & 0\end{matrix}\right)$, 
$f=x_{-\alpha}= \left(\begin{matrix} 0 & 0\\ 1 & 0\end{matrix}\right)$ and 
$h=h_{\alpha}= \left(\begin{matrix} 1 & ~0\\ 0 & -1\end{matrix}\right)$
 is a Chevalley basis for $\mathfrak{sl}_2$, with $\theta(X)=-X^T$. We have $[x_{\a},x_{-\a}]=h_{\a}$ and $\a(h_{\a})=\langle \a, h_{\a}\rangle=2$.

 Let $\mathcal{U}=\mathcal{U}(\g)$ be the universal enveloping algebra of $\g=\mathfrak{sl}_2$. Then $\mathcal{U}(\g)$ is the free associative Lie algebra generated by $e,f,h$ on which the following relations are imposed: 
$$ef-fe=h,\ hf-fh=-2f,\ he-eh=2e.$$
We have 
$$\mathfrak{sl}_2(\R)=\R e\oplus \R h\oplus \R f,$$
hence $\mathcal{U}(\g)$ contains $\mathcal{U}(\R e)$, $\mathcal{U}(\R h)$ and $\mathcal{U}(\R f)$, namely all polynomials in $e$, $h$ and $f$ as well as all their products.

 We now construct a $\Z$-form $\mathcal{U}_{\mathbb{Z}}(\g)$ of the universal enveloping algebra for $\g=\mathfrak{sl}_2(\C)$. 

  The $\Z$-form, $\mathcal{U}_{\Z}(\Mg)$ of $\mathcal{U}=\mathcal{U}(\Mg)$,  is the subring with 1 of $\mathcal{U}$ generated by
$$\left\{\dfrac{x_{\a}^m}{m!},\ \dfrac{x_{-\a}^m}{m!}\mid m\in\Z_{>0}\right\}$$
such that

\bigskip\noindent $\dfrac{(x_{\a})^m}{m!}\dfrac{(x_{-\a})^n}{n!}\  =\ \sum_{k=0}^{min(m,n)} \dfrac{(x_{-\a})^{(n-k)}}{(n-k)!} \left (\begin{matrix} h_{\a}-n-m+2k\\ k\end{matrix}\right ) \dfrac{(x_{\a})^{(m-k)}}{(m-k)!} $, 

\bigskip\noindent where if we set $u=h_{\a}-n-m+2k$ then
$ \left (\begin{matrix} u\\ k\end{matrix}\right )=\dfrac{u(u-1)(u-2)\dots (u-k+1)}{k!}$.

\bigskip\noindent This identity is due to Cartier and Kostant ([Ko]), see also ([Hu], 26.2). It follows from induction on $m$ using the following identity which holds when $m=1$:
$$x_{\a}\dfrac{(x_{-\a})^n}{n!} \ =\  \dfrac{(x_{-\a})^{n}}{(n)!}x_{\a}+ \dfrac{(x_{-\a})^{(m-1)}}{(m-1)!} (h_{\a}-m+1).$$

\begin{theorem}(Cartier-Kostant [Ko]) Let $\g=\mathfrak{sl}_2(\C)$. Let $\mathcal{U}_{\Z}(\g)$ be  the subring with 1 of $\mathcal{U}(\g)$ generated by
$\dfrac{(x_{\a})^m}{m!}$ and  $\dfrac{(x_{-\a})^m}{m!}$ with $m\in\Z_{>0}$. Then $\mathcal{U}_{\Z}(\g)$ is the free $\Z$-module with $\Z$-basis
$$\dfrac{(x_{-\a})^m}{m!} \left (\begin{matrix} h_{\a}\\ b\end{matrix}\right )\dfrac{(x_{\a})^n}{n!}$$
where $m,n,b\in\Z_{\geq 0}$.
\end{theorem}
\noindent Also, $\mathcal{U}_{\Z}(\g)$ is a $\Z$-form of $\mathcal{U}(\g)$, that is
$$\mathcal{U}_{\Z}(\g)\otimes_{\Z}\C=\mathcal{U}(\g).$$
It follows from Cartier and Kostant's Theorem that 
$\mathcal{U}(\C x_{-\a})$, $\mathcal{U}(\C h_{\a})$ and $\mathcal{U}(\C x_{\a})$ have $\Z$-subalgebras $\mathcal{U}^-_{\Z}$, $\mathcal{U}^0_{\Z}$ and $\mathcal{U}^+_{\Z}$ with $\Z$-bases consisting of $\dfrac{(x_{-\a})^m}{m!}$,  $\left (\begin{matrix} h_{\a}\\ b\end{matrix}\right )$ and $\dfrac{(x_{\a})^n}{n!}$ respectively, for $m,n,b\in\Z_{\geq 0}$ and that
$$\mathcal{U}_{\Z}=\mathcal{U}^-_{\Z}\mathcal{U}^0_{\Z}\mathcal{U}^+_{\Z}.$$

  Let $V$ be the faithful finite dimensional highest weight module with highest weight $\lambda=\omega$, the fundamental weight. Then $\lambda=\dfrac{1}{2}\alpha$ where $\alpha$ is the simple root. Let $v_{\lambda}\in V$ be a highest weight vector and let $V_{\mathbb{Z}}$ be the orbit of $v_{\lambda}$ under $\mathcal{U}_{\mathbb{Z}}$
$$V_{\mathbb{Z}}\ =\ \mathcal{U}_{\mathbb{Z}}\cdot v_{\lambda}.$$
Then 
$$\mathcal{U}^+_{\Z}v_{\lambda}=\Z v_{\lambda}$$
since all elements of the $\Z$-basis $\left\{\dfrac{(x_{\a})^n}{n!},\mid n\in\Z_{\geq 0}\right\}$ except for 1 annihilate $v_{\lambda}$. Also
$$\mathcal{U}^0_{\Z}v_{\lambda}=\Z v_{\lambda}$$
since $\left (\begin{matrix} h_{\a}\\ b\end{matrix}\right )$ acts as scalar multiplication on $v_{\lambda}$ by a $\Z$-valued scalar ([Hu], Theorem 27.1) for $b\in\Z_{\geq 0}$. Thus
$$\mathcal{U}_{\Z}\cdot v_{\lambda}=\mathcal{U}^-_{\Z}\cdot (\Z v_{\lambda})=\mathcal{U}^-_{\Z}\cdot(v_{\lambda}).$$
We have
$$\dfrac{(x_{-\a})^n}{n!}v_{\lambda}\in V_{\lambda-n\alpha},\ n\geq 0$$
where $V_{\lambda-n\alpha}$ is the weight space of $V$ of weight $\lambda-n\alpha$. Thus $\mathcal{U}^-_{\Z}$ takes $v_{\lambda}$ to weight vectors of weight lower than $\lambda$.
Set
$$V_{\R}\ =\  {\R}\otimes_{\Z} V_{\Z}.$$
Then $V_{\mathbb{Z}}$ is a lattice in $V_{\mathbb{R}}$ and
$$\dfrac{(x_{\a})^m}{m!}V_{\Z} \ \subseteq\  V_{\Z},$$
$$\dfrac{(x_{-\a})^m}{m!}V_{\Z}\ \subseteq\  V_{\Z},$$
for $m\in {\Z}_{\geq 0}$, and $i\in I$.   

We can now define the group $SL_2(\R)$ in terms of its generators as follows. We set
$$G_V(\mathbb{R})\ =\ \langle  exp(sx_{\a}), exp(tx_{-\a})\mid s,t\in\R \rangle\ =\ \langle  \left(\begin{matrix}  1 & s \\ 0 & 1  \end{matrix}\right), \left(\begin{matrix}  1 & 0 \\ t & 1  \end{matrix}\right)\mid s,t\in\R \rangle.$$
Then $$G_V(\mathbb{R}) \ \cong \ SL_2(\mathbb{\R}).$$

We have $G_V(\mathbb{R})\leq Aut(V_{\R})$. The group $SL_2(\Z)$ is defined as 
$$G_V(\mathbb{Z})=\{g\in G_V(\mathbb{R})\mid g(V_{\Z})=V_{\Z}\}.$$
Then 
$$G_V(\mathbb{Z})\ =\ \langle  exp(sx_{\a}), exp(tx_{-\a})\mid s,t\in\Z \rangle,$$
and $$G_V(\mathbb{Z}) \ \cong \ SL_2(\mathbb{\Z}).$$

It follows that every element of $SL_2(\R)$ (respectively $SL_2(\Z)$) is a product of powers of the matrices $\left(\begin{matrix}  1 & s \\ 0 & 1  \end{matrix}\right), \left(\begin{matrix}  1 & 0 \\ t & 1  \end{matrix}\right)$ and their inverses, for $s,t\in\R$ (respectively $s,t\in\Z$).

 In choosing the lattice $V_{\Z}\subset V_{\mathbb{C}}$,  the main property that is needed is that $V_{\Z}$ should be stable under the action of $\mathcal{U}_{\Z}$. The choice $V_{\Z}=\mathcal{U}_{\Z}\cdot v_\lambda=\mathcal{U}_{\Z}^-\cdot v_\lambda$, for a highest weight vector $v_\lambda$, has this property. One can also vary the choice of highest weight vector.

For example, for $SL_2$, we may choose $V_{\Z}\subset V_{\mathbb{C}}$ as follows

$$V_{\Z}=\Z\left[\begin{matrix} 1 \\ 0\end{matrix}\right]\oplus \Z\left[\begin{matrix} 0 \\ 1\end{matrix}\right],$$
$$V_{\C}=\C\left[\begin{matrix} 1 \\ 0\end{matrix}\right]\oplus \C\left[\begin{matrix} 0 \\ 1\end{matrix}\right],$$
with $$V_{\Z}=V_{\Z,\frac{1}{2}}\oplus V_{\Z,-\frac{1}{2}}$$
$$x_{-\a}:V_{\Z,\frac{1}{2}}\mapsto V_{\Z,-\frac{1}{2}}$$
where $\pm \frac{1}{2}$ denote the weights of the fundamental representation. In particular,
$$x_{-\a}v_\lambda=\left(\begin{matrix}  0 & 0 \\ 1 & 0  \end{matrix}\right)\left(\begin{matrix} 1 \\ 0 \end{matrix}\right)= 
\left(\begin{matrix} 0 \\ 1\end{matrix}\right).$$
Then $SL_2(\Z)$ is the subgroup of  $SL_2(\C)$ that stabilizes $V_{\Z}$.  Our construction above of $V_{\Z}$ in terms of orbits of a highest weight vector under $\mathcal{U}_{\Z}$ proves the existence of $V_{\Z}$ more generally, and also extends to the infinite dimensional cases.

Now let $\gamma=\left(\begin{matrix}  a & c \\ b & d  \end{matrix}\right)\in SL_2(\R)$. We make some brief remarks about the above construction in other representations. Let us first consider the adjoint representation of $SL_2(\R)$. In the basis naturally obtained from the tensor product of two fundamental representations, a general group element in the adjoint representation is given by
$$
\gamma_{\text{adj}} = \left(\begin{matrix}  a^2 & \sqrt{2} ac & c^2 \\ \sqrt{2} ab & bc+ad & \sqrt{2} cd \\ b^2 & \sqrt{2} bd & d^2 \end{matrix}\right),
$$
with the additional condition that $ad-bc=1$. 

On the other hand, we can view the integer form $SL_2(\Z)$ via the exponential  map. We first define a general group element $\gamma$ of $SL_2(\R)$ as a (finite) alternating product of its generators:
$$
\gamma = e^{x_1f}e^{x_2e}e^{x_3f}e^{x_4e}...$$
where $e,f$ are the Chevalley generators of the Lie algebra $\mathfrak{sl}_2$. In the fundamental representation, by choosing the parameters $x_i$ to be integers, the matrix representation of $\gamma$ will also contain integral entries. We next compute  the tensor product $\gamma^{(2\times 2)} = \gamma \otimes \gamma$. By projecting onto the subspace containing only the adjoint representation of $SL_2(\R)$, it can be shown that
$$
\left.\gamma^{(2\times 2)}\right\arrowvert_{\text{adj}} = e^{x_1 f_{\text{adj}}} e^{x_2 e_{\text{adj}}} e^{x_3 f_{\text{adj}}}e^{x_4 e_{\text{adj}}}...
$$
where $e_{\text{adj}}, f_{\text{adj}}$ are the matrices for the Chevalley generators $e$ and $f$ of the Lie algebra in the adjoint representation. 

We conclude  that a general element of $SL_2(\Z)$ in the adjoint representation is  generated by the exponentials of the Lie algebra generators with integer coefficients.

This argument can be generalized to other semisimple Lie groups, using the exponential map from the Lie algebra to the Lie group in their fundamental representations. 

\subsection{Integral forms of finite dimensional simple algebraic groups} We may also consider split finite dimensional simple algebraic groups in general.
Each such group has a Chevalley construction in analogy to the construction of $SL_2$ in the previous section.

 Let $\Mg$ be a finite dimensional semisimple Lie algebra over $\mathbb{C}$. Let $\mathfrak{h}$ be a Cartan subalgebra of $\Mg$ with dual  $\mathfrak{h}^{\ast}$. Let 
 $\Pi=\{\alpha_1,\dots,\alpha_{\ell}\} \subseteq
\mathfrak{h^{\ast}}$ be the simple roots and let $\Pi^{\vee}=\{\alpha_1^{\vee},\dots,\alpha_{\ell}^{\vee}\} \subseteq
\mathfrak{h}$ be the simple coroots. That is, $\langle\alpha_j,\alpha_i^{\vee}\rangle=\alpha_j(\alpha_i^{\vee})=a_{ij}$, $i,j=1,\dots \ell$, where $A=(a_{ij})_{i,j=1,\dots \ell}$ is the Cartan matrix of $\Mg$.  

Let $Q=\mathbb{Z}\alpha_1\oplus \dots \oplus \mathbb{Z}\alpha_{\ell}$ be the root lattice of $\Mg$. The {\it weight lattice} $P$ in $\mathfrak{h}^{\ast}$ is defined as follows:
$$P=\{\lambda\in \mathfrak{h}^{\ast}\mid \langle \lambda, \alpha_i^{\vee}\rangle\in\Z,\ i=1,\dots \ell\}.$$
The {\it dominant integral weights} are
$$P_+=\{\lambda\in \mathfrak{h}^{\ast}\mid  \langle \lambda, \alpha_i^{\vee}\rangle\in\Z_{\geq 0},\ i=1,\dots \ell\}.$$
The weight lattice contains a basis of {\it fundamental weights} $\{\omega_1,\dots ,\omega_{\ell}\}\subset\mathfrak{h}^{\ast}$ such that
$$\langle \omega_i, \alpha_j^{\vee}\rangle =
\begin{cases} 1,&\mbox{if } i=j\\
0,&i\neq j. \end{cases}$$
We denote the weight lattice by $P=\mathbb{Z}\omega_1\oplus \dots \oplus \mathbb{Z}\omega_{\ell}$.

 Let $\mathcal{U}=\mathcal{U}_{\mathbb{C}}(\mathfrak{g})$ be the universal enveloping algebra of $\Mg$. Let $Q^{\vee}$ be the coroot lattice, which is the $\Z$--linear span of the simple coroots
$\alpha^{\vee}_i$, for $i\in I$. Then
$$Q^{\vee}\subseteq \mathfrak{h}\subseteq \mathfrak{g}\subseteq \mathcal{U}(\mathfrak{g}).$$
Thus it makes sense to define
$$\left (\begin{matrix}
h \\ m\end{matrix}\right )= \frac{h(h-1)\dots (h-m+1)}{m!}$$ 
for $h\in\mathfrak{h}$ and $m\geq 0$.

Let ${\mathcal U}_{\mathbb{Z}}(\Mg)\subseteq {\mathcal U}_{\mathbb{C}}(\Mg)$ be the ${\mathbb{Z}}$--subalgebra generated by $\dfrac{x_{\a_i}^{m}}{m!}$,
$\dfrac{x_{-\a_i}^{m}}{m!}$ for $i\in I$ and $\left (\begin{matrix}
h\\ m\end{matrix}\right )$, for
$h\in Q^{\vee}$ and
$m\geq 0$.

 The ${\mathbb{Z}}$-subalgebra $\mathcal{U}_{\mathbb{Z}}(\mathfrak{g})$ is  a {\it ${\mathbb{Z}}$-form}  of $\mathcal{U}(\mathfrak{g})$, that is, a subring with 1 such that the 
canonical map $\mathcal{U}_{\mathbb{Z}}(\mathfrak{g})\otimes_{\mathbb{Z}}\mathbb{C}\longrightarrow \mathcal{U}(\mathfrak{g})$ is bijective. 

 Let $\lambda$ be a dominant integral weight. Let $V=V^{\lambda}$ be  the corresponding irreducible highest weight module.  For a highest weight vector $v_{\lambda}\in V^{\lambda}$, we define
$$V^{\lambda}_\mathbb{Z}\  =\ \mathcal{U}_{\mathbb{Z}}\cdot v_{\lambda}.$$
Then $V_{\Z}=V^{\lambda}_{\mathbb{Z}}$  is a $\Z$-submodule of  $V^{\lambda}_{\mathbb{R}}=V^{\lambda}_{\mathbb{Z}} \otimes_{\mathbb{Z}} \mathbb{R}$ and a ${\mathcal U}_\mathbb{Z}$-module. 

  For each weight $\mu$ of  $V^{\lambda}$, let  $V^{\lambda}_{\mu}$ be the corresponding weight
space, and we set
$$V^{\lambda}_{\mu,{\mathbb{Z}}}\ =\  V^{\lambda}_{\mu}\cap V^{\lambda}_{\mathbb{Z}}.$$
We have
$$V^{\lambda}_{\mathbb{Z}}\ =\ \oplus_{\mu\in wts(V^{\lambda})}V^{\lambda}_{\mu,{\mathbb{Z}}},$$
where the sum is taken over 
$$ wts(V^{\lambda})=\{\mu\in P\mid V^{\lambda}_{\mu}\neq 0\}.$$

Let $\mathfrak{g}_{\mathbb{R}}=\mathfrak{g}_{\mathbb{Z}} \otimes_{\mathbb{Z}} \mathbb{R}$.  Let $V=V_{\R}$ denote any faithful finite dimensional  $\Mg_{\R}$--module that has the property that the lattice generated by the weights of $V$ equals the weight lattice $P$. Here $V$ could be an irreducible highest weight module. But we also allow $V$ to be the direct sum of the fundamental modules. In this case, $V=V^{\omega_1}\oplus\dots \oplus V^{\omega_{\ell}}$ is not irreducible, but is the direct sum of irreducible highest weight modules with the fundamental weights as highest weights. We have $wts(V)=\cup_{i=1}^{\ell} wts(V^{\omega_i}),$ where for each $i\in I$
$$wts(V^{\omega_i})\subseteq \{\omega_i-\sum_{j=1}^{\ell} k_{ij}\alpha_j\mid k_{ij}\in\Z_{\geq 0}\}.$$
Hence $wts(V)$ contains all the  fundamental weights and the lattice generated by $wts(V)$ is the weight lattice $P$.

   Let $$\{x_{\a},h_i\mid \a\in\Delta,\ i\in I,\ h_i\in\mathfrak{h}\}$$
be a  Chevalley basis for $\Mg$.

   For $s,t\in \mathbb{C}$ and $i\in I$,  set 
$$\chi_{\alpha_i}(s)\ =\ exp(s\cdot x_{\a_i}),$$
$$\chi_{-\alpha_i}(t)\ =\ exp(t\cdot x_{-\a_i}).$$
Then $x_{\a_i}$ and $x_{-\a_i}$ are  nilpotent on $V$. For each $\a\in\Delta$, let 
$$\chi_{\alpha}(t)\ =\ exp(t\cdot x_{\a}).$$

  Let  $G({\mathbb{R}})=G_V(\mathbb{R})=\langle \chi_{\alpha}(t)\mid t\in\R,\ \a\in\Delta\rangle\leq Aut(V_{\mathbb{R}})$. Then $G$ is the simply connected 
{\it Chevalley group} associated to $\mathfrak{g}_{\Z}$.   If instead we chose $V$ to be the adjoint representation, this construction of $G$ would give the adjoint group.

 We define
$$G({\mathbb{Z}})=\{g\in G({\mathbb{R}})\mid g\cdot V_{\Z}= V_{\Z}\}.$$
This definition coincides for $E_7$ with the following form of $E_7(\mathbb{Z})$:
$$E_{7(+7)}(\mathbb{Z}) = E_{7(+7)}(\mathbb{R}) \cap Sp(56, \mathbb{Z})$$
discovered in [HT] following [CJ] in the framework of type II string theory. Soul\'e gave a rigorous mathematical proof that the $E_{7(+7)}(\mathbb{Z})$ of Hull and Townsend coincides with the Chevalley $\Z$--form $G(\Z)$ of $G=E_7$ ([S]). Here $E_{7(+7)}(\mathbb{R}) \cap Sp(56, \mathbb{Z})$ is the stabilizer of the standard lattice in the fundamental representation of $E_7$ which has dimension 56. The charge lattice of [HT] can be normalized to coincide with the lattice $V_{\Z}$. Once a basis for $V_{\Z}$ has been chosen, the $E_{7}(\mathbb{Z})$ orbits can be computed explicitly in terms of this basis. (See  the appendix of [MS] for a construction of the group $E_8(\Z)$ and Section~\ref{Dependence} for a discussion of the dependence of $G({\mathbb{Z}})$ on the choice of $V$).

\begin{theorem} ([CC]) Let  $\Mg$ be a simple Lie algebra and let $G$ be the corresponding Chevalley group. Let $\{\alpha_i\mid i=1,\dots , \ell\}$ be the simple roots. Then $G(\mathbb{Z})$ has the following minimal generating sets:

 (1) $\chi_{\alpha_i}(1)$ and $\chi_{-\alpha_i}(1)$, $i=1,\dots , \ell$, 

 or 

 (2) $\chi_{\alpha_i}(1)$ and $\widetilde{w}_{\alpha_i}=\chi_{\alpha_i}(1)\chi_{-\alpha_i}(-1)\chi_{\alpha_i}(1)$, $i=1,\dots , \ell$.

\end{theorem}
 {The generating set (2) is the analog of the $S,T$-generating set for $SL_2(\Z)$}.

\subsection{Construction of Kac--Moody groups  over $\R$ and $\Z$}

  Now let $\Mg$ be a Kac--Moody algebra.  We wish to construct the Kac--Moody group $G$ in analogy with the Chevalley construction of finite dimensional semisimple algebraic groups in the previous section. The generalization of this method to Kac--Moody groups is provided by [CG2].  With an integrable highest weight module $V$ for $\Mg$ and a $\mathbb{Z}$-form $V_{\mathbb{Z}}$, we are able to construct our Kac--Moody group over $\R$ and over $\mathbb{Z}$.

\subsubsection{Roots and highest weight modules}
Let $\Mg$ be a Kac--Moody algebra. Let $\mathfrak{h}$ be a Cartan subalgebra.   Every root $\alpha\in\Delta$ has an expression in $Q$ of the form $\alpha=\sum_{i=1}^{\ell} k_i\alpha_i$ where the $k_i$ are either all $\geq 0$, in which case $\alpha$ is called {\it positive}, or all $\leq 0$, in which case $\alpha$ is called {\it negative}. The positive roots are denoted $\Delta_+$, the negative roots $\Delta_-$.  A root $\alpha\in\Delta$ is called a {\it real root} if there exists $w\in W$ such that $w\alpha$ is a
simple root. A root $\alpha$ which is not real is called {\it
imaginary}. We denote by $\Delta^{re}$ the real roots and  $\Delta^{im}$ the
imaginary roots.

We recall that a $\mathfrak{g}$--module $V$ is called {\it integrable} if it is diagonalizable, that is, $V$ can be written as a direct sum of its weight spaces, and the $e_i$ and $f_i$ act locally nilpotently on $V$. That is, for each $v\in V$, $e_i^n\cdot v=f_i^n\cdot v=0$ for all $i$ and for all $n>>0$.

Let $\mathfrak{g}^+ =\bigoplus_{\alpha\in\Delta_+}\mathfrak{g}_{\alpha}$. Let $V$ be a  representation of $\Mg$. Then $V$ is called a {\it highest weight representation} with highest weight $\lambda\in\mathfrak{h}^{\ast}$ if there exists $0\neq v_{\lambda}\in V$ such that
$$\mathfrak{g}^{+}(v_{\lambda})=0,$$
$$h(v_{\lambda})={\lambda}(h)v_{\lambda}$$
for $h\in\mathfrak{h}$ and
$$V=\mathcal {U}(\mathfrak{g})(v_{\lambda}).$$
Since $\mathfrak{g}^{+}$ annihilates $v_{\lambda}$, $\mathfrak{h}$ acts as scalar multiplication on $v_{\lambda}$, we have
$$V=\mathcal {U}(\mathfrak{g}^-)(v_{\lambda}).$$
There is a unique irreducible highest weight module $V=V^{\lambda}$ for each dominant integral weight $\lambda$. 

\subsubsection{$\Z$-form of $\mathcal{U}(\g)$ for $\g$ a Kac--Moody algebra.}

Let ${\mathcal U}$, ${\mathcal U}^+$ and ${\mathcal U}^-$ be the universal enveloping algebras of 
$\mathfrak{g}$, $\mathfrak{g}^{+}$ and $\mathfrak{g}^{-}$ respectively. 

\medskip \noindent Let $V$ be  a highest weight module with highest weight $\lambda\in\mathfrak{h}^{\ast}$. For a highest weight vector $v_{\lambda}\in V$, we have
$$\mathfrak{g}^{+}(v_{\lambda})=0,\ h(v_{\lambda})={\lambda}(h)v_{\lambda}$$
for $h\in\mathfrak{h}$ and
$$V=\mathcal {U}(\g)(v_{\lambda}).$$
By the Poincar\'e-Birkhoff-Witt theorem, $\mathcal {U}(\g)=\mathcal {U}(\g)^-\otimes \mathcal {U}(\mathfrak{h}) \otimes\mathcal {U}(\g^+)$.
Since $\mathfrak{g}^{+}$ annihilates $v_{\lambda}$ and $\mathfrak{h}$ acts as scalar multiplication on $v_{\lambda}$,  we have
$$V=\mathcal {U}(\g^-)(v_{\lambda}).$$

\noindent We let $\Lambda\subseteq \mathfrak{h}^*$ be the $\Z$--linear span of  the simple roots $\alpha_i$, for $i\in I$, and $\Lambda^{\vee}\subseteq
\mathfrak{h}$ be the $\Z$--linear span of the simple coroots
$\a^{\vee}_i$, for $i\in I$.
Let 

 ${\mathcal U}^+_{{\Z}}\subseteq {\mathcal U}^+_{{\C}}$ be the ${\Z}$-subalgebra generated by $\dfrac{e_i^{m}}{m!}$ for $i\in I$
and
$m\geq 0$,  

 ${\mathcal U}^-_{{\Z}}\subseteq {\mathcal U}^-_{{\C}}$ be the ${\Z}$-subalgebra generated by $\dfrac{f_i^{m}}{m!}$ for $i\in I$
and
$m\geq 0$,  

 ${\mathcal U}^0_{{\Z}}\subseteq {\mathcal S}(\mathfrak{h}_{\C})$ be the ${\Z}$-subalgebra generated by $\left (\begin{matrix}
h \\ m\end{matrix}\right )$, for
$h\in\Lambda^{\vee}$ and $m\geq 0$, where $\mathcal{S}$ is the symmetric algebra of $\mathfrak{h}_{\C}$,

 ${\mathcal U}_{{\Z}}\subseteq {\mathcal U}_{{\C}}$ be the ${\Z}$-subalgebra generated by $\dfrac{e_i^{m}}{m!}$,
$\dfrac{f_i^{m}}{m!}$ for $i\in I$ and $\left (\begin{matrix}
h\\ m\end{matrix}\right )$, for
$h\in\Lambda^{\vee}$ and
$m\geq 0$.
 
  By a {\it ${\Z}$-form} of $\mathcal{U}(\g)$, we mean a subring $\mathcal{U}_{\Z}(\g)$ such that the canonical map $$\mathcal{U}_{\Z}(\g)\otimes_{\Z}\C\longrightarrow \mathcal{U}(\g)$$ is bijective.  It follows ([Ti]) that ${\mathcal U}_{\Z}$ is a ${\Z}$-form of ${\mathcal U}_{\C}$.

\subsubsection{$\Z$-form of a highest weight module}\label{Zform}

  We shall construct a lattice $V_{\Z}$ in $V$ by taking the orbit of a highest weight vector $v_{\lambda}$ under $\mathcal{U}_{\Z}$. We have
$$\mathcal{U}^+_{\Z}v_{\lambda}=\Z v_{\lambda}$$
since all elements of ${\mathcal U}^+_{{\Z}}$ except for 1 annihilate $v_{\lambda}$. Also
$$\mathcal{U}^0_{\Z}v_{\lambda}=\Z v_{\lambda}$$
since ${\mathcal U}^0_{{\Z}}$ acts as scalar multiplication on $v_{\lambda}$ by a $\Z$-valued scalar. Namely, $\left (\begin{matrix}
h\\ m\end{matrix}\right )$, for
$h\in\Lambda^{\vee}$ and $m\geq 0$ acts on $v_{\lambda}$ as
$$\left (\begin{matrix}\lambda(h)\\ m\end{matrix}\right )=\dfrac{\lambda(h)(\lambda(h)-1)\dots (\lambda(h)-m+1)}{m!}\in\Z.$$
Thus
$$\mathcal{U}_{\Z}\cdot v_{\lambda}=\mathcal{U}^-_{\Z}\cdot (\Z v_{\lambda})=\mathcal{U}^-_{\Z}\cdot(v_{\lambda}).$$
 Let $\alpha$ be any real root  of $\mathfrak{g}$ and let $e_{\alpha}$ and $f_{\alpha}$ be root vectors corresponding to $\alpha$. Let $V$ be an integrable  highest weight module with highest weight $\lambda$ and highest weight vector $v_{\lambda}$. Then 

$$f_{\alpha} v_{\lambda}=v_{\lambda-\alpha},$$

$$\dfrac{f_{\alpha}^m}{m!}v_{\lambda}\in V_{\lambda-m\alpha}.$$
For a weight $\mu<\lambda$ we have

$$e_{\alpha} v_{\mu}=v_{\mu+\alpha},$$

$$\dfrac{e_{\alpha}^m}{m!}v_{\mu}\in V_{\mu+m\alpha}.$$
In particular, 

$$\dfrac{e_{\alpha}^m}{m!}v_{\lambda-m\alpha}=c_{\lambda} v_{\lambda}$$
for some constant $c_{\lambda}$.   We set 
$$V_\Z\  =\ \mathcal{U}_{\Z}\cdot v_{\lambda}\ =\ \mathcal{U}^-_{\Z}\cdot(v_{\lambda})$$
Then $V_{\Z}$  is a lattice in  $V_{\C}$ and a ${\mathcal U}_\Z$-module.

  For each weight $\mu$ of  $V$, let  $V_{\mu}$ be the corresponding weight
space, and we set
$$V_{\mu,{\Z}}\ =\  V_{\mu}\cap V_{\Z}.$$
We have then
$$V_{\Z}\ =\ \oplus_{\mu}V_{\mu,{\Z}},$$
where the sum is taken over the weights of $V$. Thus $V_{\Z}$ is a direct sum of its weight spaces. We set
$$V_{\mu,\R}\ =\ \R\otimes_{\Z} V_{\mu,\Z}$$
so that 
$$V_{\R}\ =\ \R\otimes_{\Z} V_{\Z}\ =\ \oplus_{\mu}V_{\mu,{\R}}.$$

  For each weight
$\mu$ of $V$, $\mu=\lambda-\sum_{i=1}^{\ell} k_i\alpha_i$, where $\lambda$ is the highest weight and  $k_i\in{\Z}_{\geq 0}$. Define the {\it depth} of
$\mu$ to be
$$depth(\mu)\quad=\quad \sum_{i=1}^{\ell} k_i.$$
A basis $\Psi=\{v_1,v_2,\dots\}$ of $V$ is called {\it coherently ordered relative to depth} if

 (1) $\Psi$ consists of weight vectors.

 (2) If $v_i\in V_{\mu}$, $v_j\in V_{\mu'}$ and $depth(\mu')\ >\ depth(\mu)$, then $j>i$.

 (3) $\Psi\cap V_{\mu}$ consists of an interval $v_k,v_{k+1},\dots,v_{k+m}$.

\begin{theorem} ([G1]) The lattice $V_{\Z}$ has a coherently ordered $\Z$-basis $\{v_1,v_2,\dots\}$ where $v_i\in V_{\Z}$,  $v_i=\xi v_{\lambda}$, $\xi\in \mathcal{U}_{\Z}$, $w_i=k_i\otimes v_i$, $k_i\in \R-\{0\}$ and $\{w_1,w_2,\dots \}$ is a  coherently ordered basis for $V_{\R}$. Vectors in $V_{\Z}$  have integer valued norms relative to a Hermitian inner product $\{\cdot ,\cdot\}$ on $V$.\end{theorem}

  To construct the inner product $\{\cdot ,\cdot\}$  on $V_{\Z}$, we take the Cartan involution $\ast$ on $\mathfrak{g}$, which is a conjugate--linear automorphism on $\mathfrak{g}$, taking $\mathfrak{g}_{\alpha}$ to $\mathfrak{g}_{-\alpha}$ and preserving a `compact' real form. We use this to define a conjugate--linear anti--automorphism $\ast$ on the tensor algebra and we extend this to the universal enveloping algebra $\mathcal{U}$.  We then define  a positive definite Hermitian inner product $\{\cdot ,\cdot\}$  on $V^{\lambda}=\mathcal{U}\cdot v_{\lambda}$ in terms of the projection of $\ast$ on $\mathcal{U}^0=\mathcal{U}(\mathfrak{h})$ and show that this is well defined on $V^{\lambda}_{\Z}$. Relative to $\{\cdot ,\cdot\}$, all basis vectors  have $\Z$-valued norm, and $\{ v_{\lambda}, v_{\lambda}\}=1$.

\subsubsection{The Kac--Moody group $G(\R)$}

 Our next step is to construct our Kac--Moody group $G(\R)$. Let $V$ be  an integrable highest weight module for $\mathfrak{g}$. We note that  $e_i$
and $f_i$ are locally nilpotent on  $V$.

 We let $V_{\Z}$ be a $\Z$-form of $V$ as in Subsection~\ref{Zform}. Since $V_{\Z}$ is a $\mathcal{U}_{\Z}$-module, we have
$$\dfrac{e_i^m}{m!} (V_{\Z})\subseteq V_{\Z},$$
$$\dfrac{f_i^m}{m!} (V_{\Z})\subseteq V_{\Z},$$
for $m\in {\Z}_{\geq 0}$, and $i\in I$.

  For $s,t\in \R$ and $i\in I$,  set 
$$\chi_{\alpha_i}(s)\ =\ \sum_{m\in {\Z}_{\geq 0}} s^m \dfrac{e_i^m}{m!}\ =\ exp(se_i),$$
$$\chi_{-\alpha_i}(t)\ =\ \sum_{m\in {\Z}_{\geq 0}} t^m \dfrac{f_i^m}{m!}\ =\ exp(tf_i).$$
Then $\chi _{\alpha _{i}}(s),$ $\chi
_{-\alpha
_{i}}(t)$ define elements in $Aut(V_{\R})$, thanks to the
local
nilpotence of $e_{i},$ $f_{i}.$  

  Relative to a coherently ordered
basis of $V_{\Z}$, the elements 
$\chi_{\alpha}(u)$, $u\in \R$, $\alpha$ a positive real root, are represented by infinite upper triangular matrices with 1's on the diagonal, and the elements $\chi_{\alpha}(u)$,
$u\in \R$, $\alpha$ a negative real root, are represented by infinite lower triangular matrices with 1's on the diagonal.

  We let $G(\R)$ be the subgroup of $Aut(V_{\R})$ generated by the linear automorphisms $\chi_{\alpha_i}(s)$ and  $\chi_{-\alpha_i}(t)$ of
$V_{\R}$, for $s,t\in \R$. That is, 
$$G(\R)=\langle exp(se_i), exp(tf_i)\mid  s,t\in \R\rangle.$$

\subsubsection{The $\Z$-form $G(\Z)$}

 As in [CG2], we define the `$\Z$-form' $G(\mathbb{Z})$ of $G(\mathbb{R})$ in the following way. We set
$$G(\mathbb{Z})=G(\mathbb{R})\cap Aut(V_{\mathbb{Z}}).$$
Then $$G(\mathbb{Z})=\{\gamma\in G(\R)\mid \gamma\cdot V_{\mathbb{Z}}= V_{\mathbb{Z}}\}.$$

\subsubsection{Dependence of $G$ on the choice of representation $V$ and the lattice $V_{\Z}$}\label{Dependence}

 Let $\Mg$ be a Kac--Moody algebra. Let $\mathcal{S}$ be the set of integrable highest weight modules for $\Mg$ whose sets of weights contains all the fundamental weights. 
 
 If $V_1$ and $V_2$ belong to $\mathcal{S}$, then the Kac--Moody groups $G_{V_1}$ and $G_{V_2}$ are isomorphic when constructed over $\R$ or $\C$. However, this may no longer be true when $G_{V_1}$ and $G_{V_2}$ are constructed over $\Z$. In this case (over $\Z$),  $G_{V_1}$ and $G_{V_2}$ depend on the choice of representations $V_1$ and $V_2$ ([CG2]).

The construction of $G$ over $\R$ and $\Z$ does depend on the choice of lattice $\Lambda$ of weights of the representation $V$. In the affine case, Garland was able to completely characterize the dependence by constructing an appropriate cocycle ([G5]).
 
 In general, we have $Q\leq\Lambda\leq P$ where $P$ is the weight lattice and the $Q$ is the root lattice. In most of our applications, $Q=P$ or $Q$ has index at most 2 in $P$. The group $G$ corresponding to $Q$ is the adjoint group (corresponding to the adjoint representation) and the group $G$ corresponding to $P$ is the simply connected group. In our applications, we will usually construct the simply connected group. When $Q=P$, the adjoint and simply connected groups coincide.

\subsubsection{Varying the choice of $V$ and of $V_{\Z}$}
As we saw in Subsection~\ref{Dependence}, given an integrable highest weight module $V$, we are free to choose an admissible lattice $V_{\Z}$. Such a lattice $V_{\Z}$ is $\mathcal{U}_{\Z}$--invariant and contains a highest weight vector $v_{\lambda}$. In Section~\ref{SL2Chev}, for  $SL_2$ in the defining representation, we chose $V_{\Z}=\Z\oplus \Z$ and this choice of  $V_{\Z}$ has all the required properties. However, in the finite dimensional case, choosing $V_{\Z}=\Z\oplus\dots\oplus \Z$ ($n$--factors)  does not have the required properties in general. Though we can make this choice for $V_{\Z}$ whenever $V$ is the standard (defining) matrix representation. Some examples are the $n\times n$ defining representation of $SL_n$ and the the $2n\times 2n$ defining representation of $SO(n,n)$. These examples have the property that the standard set of basis vectors is an orthonormal basis for $V_{\Z}$ and $V_{\Z}$ is spanned by the $\Z$--span of this set of basis vectors.

 {\bf Example} Let $G=SL_2$. In the standard representation $V_{\C}=\C\oplus\C$ with standard basis 
$x_{\alpha}= \left(\begin{matrix} 0 & 1\\ 0 & 0\end{matrix}\right)$, 
$x_{-\alpha}= \left(\begin{matrix} 0 & 0\\ 1 & 0\end{matrix}\right)$ and 
$h_{\alpha}= \left(\begin{matrix} 1 & ~0\\ 0 & -1\end{matrix}\right)$, we have $V_{\Z}=\Z\oplus\Z$. Then $V_{\Z}$ has stabilizer (in the defining representation of $\mathfrak{sl}_2$)
$$\Z x_{-\a}+\Z h_{\a}+ \Z x_{\a}.$$
However, the stabilizer of $V_{\Z}$ under the adjoint representation of $\mathfrak{sl}_2$ is
$$\Z x_{-\a}+\Z\left(\frac{h_{\a}}{2}\right)+ \Z x_{\a}.$$
These do not give rise to the same form of the discrete group $G(\Z)$. In the defining representation, $G(\Z)=SL_2(\Z)$. In the adjoint representation, $G(\Z)=PSL_2(\Z)$. However, when the root lattice and weight lattice coincide (as is the case for $E_8$ and $E_{10}$), the defining representation and the adjoint representation give the same form of group $G(\Z)$.  In general, the number of distinct forms of $G(\Z)$ depends on the index of the root lattice as a subgroup of the weight lattice (which equals 2 for $SL_2$). All lattices $\Lambda$ that lie between the root lattice $Q$ and the weight lattice $P$ can be realized as lattices of weights for some faithful representation $V$. Each such lattice $\Lambda$ gives rise to a distinct form of $G(\Z)$.

\subsubsection{Finding $G(\Z)$--orbits on $V_{\Z}$}

 We have  $V_{\mathbb{Z}}= \mathcal{U}_{\mathbb{Z}}\cdot v_\lambda$, where $v_\lambda$ is a highest weight vector.  Let $\alpha$ be a positive real  root and let $\leq$ be a partial order on the weight lattice $P$. Let $\mu<\lambda$ be a weight lower than $\lambda$. We have
$$\chi_{-\alpha}(t)\cdot v_{\lambda}=\sum_{ \mu<\lambda} t^{c(\mu)}v_{\mu},$$
$$\chi_{\alpha}(s)\cdot v_{\mu}=\sum_{ \mu<\rho\leq \lambda} s^{c(\rho)}v_{\rho}.$$
Thus if $\langle \lambda,\ \alpha\rangle\neq 0$, then 
$\chi_{-\alpha}(t)\cdot v_{\lambda}\in V_{\mathbb{Z}}$ implies that $t^{c(\mu)}\in\Z$, for $c(\mu)\in\Z_{\geq 0}$ so $t\in\Z$. Similarly,
$\chi_{\alpha}(s)\cdot v_{\mu}\in V_{\mathbb{Z}}$ implies that  $s^{c(\rho)}\in\Z$, for $c(\rho)\in\Z_{\geq 0}$ so $s\in\Z$. 

 Since the $\chi_{-\alpha}(t)$ and $\chi_{\alpha}(s)$, for $s,t\in\Z$, generate $G(\Z)$, the above formulas can be used to compute the orbits of $G(\Z)$ on $V_{\mathbb{Z}}$, after choosing a coherently ordered basis for $V_{\mathbb{Z}}$.

 We note that $\chi_{-\alpha}(t)\cdot v_{\lambda}$ is a $\Z$--linear sum of weight vectors of weights {\it lower} than $\lambda$ and for $\mu<\lambda$, $\chi_{\alpha}(s)\cdot v_{\mu}$ is a $\Z$--linear sum of weight vectors of weights {\it higher} than $\mu$ but $\leq \lambda$.

 {\bf Example} Consider $SL_2$ in the standard representation with fundamental coroot \linebreak $h_{\alpha}=\alpha^{\vee}= \left(\begin{matrix} 1 & ~0\\ 0 & -1\end{matrix}\right)$. The fundamental weights are $-\omega$, $0$, $\omega$ with $\omega(\alpha^{\vee})=1$. We may choose $V_{\Z}\subset V_{\mathbb{C}}$ as follows
$$V_{\Z}=\Z\left(\begin{matrix} 1 \\ 0\end{matrix}\right)\oplus \Z\left(\begin{matrix} 0 \\ 1\end{matrix}\right),$$
$$V_{\C}=\C\left(\begin{matrix} 1 \\ 0\end{matrix}\right)\oplus \C\left(\begin{matrix} 0 \\ 1\end{matrix}\right),$$
with $$V_{\Z}=V_{\Z,\omega}\oplus V_{\Z,-\omega}.$$
We have
$$x_{-\alpha}\cdot v_\lambda=\left(\begin{matrix}  0 & 0 \\ 1 & 0  \end{matrix}\right)\left(\begin{matrix} 1 \\ 0 \end{matrix}\right)= 
\left(\begin{matrix} 0 \\ 1\end{matrix}\right)$$
so
$$x_{-\alpha}:V_{\Z,\omega}\mapsto V_{\Z,-\omega}$$
and
$$x_{\alpha}\cdot v_{-\lambda}=\left(\begin{matrix}  0 & 1 \\ 0 & 0  \end{matrix}\right)\left(\begin{matrix} 0 \\ 1 \end{matrix}\right)= 
\left(\begin{matrix} 1 \\ 0\end{matrix}\right)$$
so
$$x_{\alpha}:V_{\Z,-\omega}\mapsto V_{\Z,\omega}.$$
Also
$$h_{\alpha}\cdot v_\lambda=\left(\begin{matrix}  1 & ~0 \\ 0 & -1  \end{matrix}\right)\left(\begin{matrix} 1 \\ 0 \end{matrix}\right)= 
\left(\begin{matrix} 1 \\ 0\end{matrix}\right)$$
It follows that the $SL_2(\Z)$--orbit on $V_{\Z}$ is contained  in a set of vectors of the form $$\left\{\left(\begin{matrix} m \\ n\end{matrix}\right)\mid m,n\in\Z\right\}$$ where
$$\left(\begin{matrix} m \\ n\end{matrix}\right)=m\left(\begin{matrix} 1 \\ 0\end{matrix}\right)+n\left(\begin{matrix} 0 \\ 1\end{matrix}\right)=mv_{\lambda}+nv_{-\lambda}.$$

  In general we have
$$V_{\mathbb{Z}}\ =\ \oplus_{\mu\in wts(V)}V_{\mathbb{Z},\mu},$$
where the sum is taken over 
$$ wts(V)=\{\mu\in P\mid V_{\mu}\neq 0\}.$$




\subsubsection{Iwasawa decomposition of $G$}
The group $G(\R)$ has unique Iwasawa decomposition ([KP], [DGH])
$$G(\R) = KA^+N.$$
The subgroup $K$ is the fixed point subgroup of the  involution on $G(\R)$ induced from the Cartan involution on the Lie algebra $\mathfrak{g}_{\R}$ and is the analog of the maximal compact  subgroup. Let $A^+$ be the analog of the diagonal subgroup with positive entries. Let $N$ be the subgroup generated by all positive real root groups. Then $N$ is the analog of the upper triangular subgroup with 1's on the diagonal.

\subsection{Generating sets for $G(\Z)$}

 The following lemma  shows that  for a symmetrizable Kac--Moody group $G$, taking `$\Z$-points'  in the generators of $G(\R)$ preserves $V_{\Z}$.
\begin{lemma} ([CG2])  Let $s,t\in\Z$. Then 
$$\chi_{\alpha_i}(s)(V_{\Z})=exp(se_i)(V_{\Z})\subseteq V_{\Z},$$
$$\chi_{-\alpha_i}(t)(V_{\Z})=exp(tf_i)(V_{\Z})\subseteq V_{\Z}.$$
\end{lemma}

In  [CG2], the authors conjecture the stronger statement that the group generated by $\chi_{\alpha}(s)$, for $\alpha\in\Delta^{re}$ and  $s\in\Z$, equals $G(\Z)$. Since indexing over the full set of real roots would give redundant generators, a corollary of this would be the following.

\begin{corollary} ([CG2]) Let $G$ be a Kac--Moody group corresponding to a symmetrizable generalized Cartan matrix. Let  $\{\alpha_i\mid i=1,\dots , \ell\}$ be the simple roots. Then $G(\mathbb{Z})$ has the following finite minimal generating sets:

 (1) $exp(e_i)$ and $exp(f_i)$, $i=1,\dots , \ell$, 

 or 

 (2) $exp(e_i)$ and $\widetilde{w}_{\alpha_i}=exp(e_i)exp(-f_i)exp(e_{-i})$, $i=1,\dots , \ell$.

\end{corollary}

The conjecture of [CG2] is confirmed in [AC] where a {\it finite presentation} of $G(\mathbb{Z})$ is given.

\subsection{Root lattice and weight lattice}

Let $Q$ be the root lattice and let $P$ be the weight lattice. As in the finite dimensional case, the index of $Q$ in $P$ is finite, and is given by $|det(A)|$ where $A$ is the corresponding generalized Cartan matrix ([CS]). The following formula may be proven recursively.

\begin{lemma} Let $A$ be the generalized Cartan matrix of a Dynkin diagram of the form $Y(p,q,r)$ (which has $p+q+r-2$ nodes). Then $det(A)= -pq-qr-rp+pqr$.\end{lemma}

\medskip\noindent The Dynkin diagram for $E_{10}$ is of the form $Y(2,3,7)$. Thus $|det(A(E_{10}))|=1$, so the root lattice and weight lattice coincide. The Dynkin diagram for $E_{11}$ is of the form $Y(2,3,8)$. Thus  $|det(A(E_{11}))|=2$, so the root lattice $Q$ has index 2 in the weight lattice $P$.

\subsection{Explicit construction of the groups $E_9$, $E_{10}$ and $E_{11}$} 

 We summarize the constructions of the groups $E_{n}(\mathbb{R})$ and $E_{n}(\mathbb{Z})$ for $n=9,10,11$. In each of these constructions, we take the following data (which includes a particular choice of representation $V$ for the Kac--Moody algebra  $\mathfrak{e}_{n}$), together with the data in Table~\ref{data}:

\medskip

 $\mathcal{U}_{\mathbb{Z}}$, a $\Z$-form of the universal enveloping algebra $\mathcal{U}_{\mathbb{C}}$ 

$V_{\mathbb{Z}}=\mathcal{U}_{\mathbb{Z}}^-\cdot v_\lambda$, where $v_\lambda$ is a highest weight vector of $V$

 $\mathcal{U}_{\mathbb{Z}}^-=\mathbb{Z}$-subalgebra of $\mathcal{U}_{\mathbb{C}}$ generated by $\dfrac{f_i^m}{m!}$, $i=1,\dots ,n$

\begin{table}[ht!]
\caption{}
\centering
\begin{tabular}{| c | c |  c  | c | c|}
\hline
Algebra & Generators  & Simple roots & Fund. weights & Highest wt. module\\
\hline

   $\mathfrak{e}_{9}(\mathbb{C})$
& 
$\begin{matrix}
& e_1,\dots ,e_{9}\\
& f_1,\dots ,f_{9}
\end{matrix}$

&
$
 \alpha_1,\dots ,\alpha_{9}
$

&
 $\omega_1,\dots ,\omega_{9}$
 
&  $\begin{matrix}
\text{$V=V^{\omega_{1}}$  }\\ 
\text{$V$ integrable  with high. wt. vec. $v^{\omega_1}$ }\\
\text{corresp. to fund. weight $\omega_{1}$ }
 \text{labeled as in}\\
  \text{Figure~\ref{fig:DynkinE11}}
\end{matrix}$

\\ \hline

   $\mathfrak{e}_{10}(\mathbb{C})$
& 
$\begin{matrix}
& e_1,\dots ,e_{10}\\
& f_1,\dots ,f_{10}
\end{matrix}$

&
$
 \alpha_1,\dots ,\alpha_{10}$

&
 $\omega_1,\dots ,\omega_{10}$
 
&  $\begin{matrix}
\text{$V=V^{\omega_1+\dots +\omega_{10}}$  }\\ 
\text{$V$ irred. and integrable with high. wt. vec. }\\
v^{\omega_1+\dots +\omega_{10}} \\
\end{matrix}$

\\ \hline

   $\mathfrak{e}_{11}(\mathbb{C})$
& 
$\begin{matrix}
& e_1,\dots ,e_{11}\\
& f_1,\dots ,f_{11}
\end{matrix}$

&
$\begin{matrix}
 \alpha_1,\dots ,\alpha_{11}\\
 \text{labeled as in}\\
  \text{Figure~\ref{fig:DynkinE11}}
\end{matrix}$

&
 $\omega_1,\dots ,\omega_{11}$
 
&  $\begin{matrix}
\text{$V=\omega_{11}$--rep. of $\mathfrak{e}_{11}(\mathbb{C})$.  }\\ 
\text{$V$ integrable  with high. wt. vec. $v^{\omega_1}$ }\\
\text{corresp. to fund. weight $\omega_{11}$ }
\end{matrix}$

\\ \hline
\end{tabular}
\label{data}
\end{table}

With this data, we have
 
 $$\boxed{E_{n}(\mathbb{R})=\langle  \chi_{\alpha_i}(s)=exp(se_i)$, $\chi_{-\alpha_i}(t)=exp(tf_i) \mid s,t\in\mathbb{R}\ i=1,\dots ,n\rangle}$$

 $$\boxed{E_{n}(\mathbb{Z})=E_{n}(\mathbb{R})\cap Aut(V_{\mathbb{Z}})=\langle  \chi_{\alpha_i}(s)=exp(se_i)$, $\chi_{-\alpha_i}(t)=exp(tf_i) \mid s,t\in\mathbb{Z},\ i=1,\dots ,n\rangle}$$

\medskip
We now describe in more detail the choice of module $V$ for the construction of our Kac--Moody groups.  We refer the reader to Subsection~\ref{Dependence} for statements as to how $E_{n}(\mathbb{R})$ and $E_{n}(\mathbb{Z})$ depend on the choice of $V$.

\medskip
{\it {$\circ$ $E_{9}(\mathbb{Z})$ and $E_{9}(\mathbb{R})$} }

For $E_9$, we choose $V$ to be the fundamental representation $V^{\omega_1}$ which is the highest weight module with highest weight $\omega_1$, where $\omega_1$ is the fundamental weight dual to $\alpha_1$ as labeled in Figure~\ref{fig:DynkinEn}. Our motivation for this is the following. As we discuss in the appendix, the coefficients in front of the $\mathcal{R}^4$ and $\partial^4\mathcal{R}^4$ correction terms are the functions $E_{(0,0)}$ and $E_{(1,0)}$. These functions are related to Eisenstein series on various finite dimensional groups. In particular $E_{(0,0)}$ is an Eisenstein series characterized by a scalar multiple of $\omega_1$. Therefore, if we construct Eisenstein series using representation theory, it is convenient to use the fundamental representation $V^{\omega_1}$ . For $A_1$, $E_6$, $E_7$ and $E_8$, $E_{(1,0)}$ also corresponds to the representation $V^{\omega_1}$.

\bigskip

{\it {$\circ$ $E_{10}(\mathbb{Z})$ and $E_{10}(\mathbb{R})$}}

When the root lattice $Q$ equals the weight lattice $P$, as is the case for $\mathfrak{e}_{10}$, that is, $|det(A)|=1$, where $A$ is the generalized Cartan matrix, then we may choose 
$V=V^{\omega_1+\dots +\omega_{10}}$, where $\omega_i$ are the fundamental weights. Then $V$ is an irreducible integrable highest weight module with lattice of weights equal to $P$ and with highest weight $\omega_1+\dots +\omega_{10}$. We have thus
$$wts(V)\subseteq \{\omega_1+\dots +\omega_{10}-\sum_{j=1}^{10} k_{j}\alpha_j\mid k_{j}\in\mathbb{Z}_{\geq 0}\},$$
where $\alpha_i$ are the simple roots, and
$$\left(\begin{matrix}
\omega_1 \\
\omega_2\\
\vdots\\
\omega_{10}
\end{matrix}\right)=A^{-1}
\left(\begin{matrix}
\alpha_1 \\
\alpha_2\\
\vdots\\
\alpha_{10}
\end{matrix}\right).$$
Since the root lattice $Q=\mathbb{Z}\alpha_1\oplus \dots \oplus \mathbb{Z}\alpha_{10}$ and the weight lattice $P$ coincide, the weights of $V$ are contained in the $\mathbb{Z}$-span of the simple roots. Hence $wts(V)$ contains all the fundamental weights $\omega_i$. 

\bigskip
{\it{$\circ$ $E_{11}(\mathbb{Z})$ and $E_{11}(\mathbb{R})$}}

 Symmetries of the discrete group $E_{11}(\mathbb{Z})$ were discussed in [GW1], where the authors conjecture that the group $E_{11}(\mathbb{Z})$ preserves the brane charge lattice. This charge lattice belongs to the fundamental representation $V$ corresponding to the vertex at the end of the long tail of the Dynkin diagram for $E_{11}$. We call this representation  the $\omega_{11}$--representation. In [GW1], due to a different labeling of the vertices of the Dynkin diagram, this is called the $\ell_1$--representation.

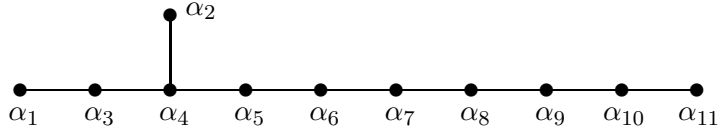
\begin{figure}
\begin{center}
\setlength{\unitlength}{1mm}
\begin{picture}(90,10)
   \put(0,0){\circle*{1.7}}
   \put(10,0){\circle*{1.7}}
   \put(20,0){\circle*{1.7}}
   \put(30,0){\circle*{1.7}}
   \put(40,0){\circle*{1.7}}
   \put(50,0){\circle*{1.7}}
   \put(60,0){\circle*{1.7}}
   \put(70,0){\circle*{1.7}}
   \put(80,0){\circle*{1.7}}
   \put(90,0){\circle*{1.7}}
   \put(20,10){\circle*{1.7}}

   \put(0,0){\line(1,0){90}}
   \put(20,0){\line(0,1){10}}

   \put(-1.5,-4){$\alpha_1$}
   \put(8.5,-4){$\alpha_3$}
   \put(22,10){$\alpha_2$}
   \put(18.5,-4){$\alpha_4$}
   \put(28.5,-4){$\alpha_5$}
   \put(38.5,-4){$\alpha_6$}
   \put(48.5,-4){$\alpha_7$}
   \put(58.5,-4){$\alpha_8$}
   \put(68.5,-4){$\alpha_9$}
   \put(77.5,-4){$\alpha_{10}$}
   \put(87.5,-4){$\alpha_{11}$}
 
\end{picture}

\end{center}
\bigskip
  \caption{The Dynkin diagram of $E_{11}$}
  \label{fig:DynkinE11}
\end{figure}

 Our general construction gives $E_{11}(\mathbb{Z})$ as the subgroup of $E_{11}(\mathbb{R})$ that preserves a lattice in the $\omega_{11}$--representation of the Kac--Moody algebra $\mathfrak{e}_{11}(\R)$. Our construction depends on the choice of representation and differs from the construction in [CG2] where the tensor product of the fundamental representations was used. Here our choice of representation is motivated by the conjecture of Gubay and West ([GW1]) that $E_{11}(\mathbb{Z})$ preserves the brane charge lattice based on this representation.

\subsection{Structure of $K(G(\R))\backslash G(\R)/G(\Z)$}

It would be advantageous to obtain  a fundamental domain for the action of  $G(\mathbb{Z})$ on $K(G(\mathbb{R}))\backslash G(\mathbb{R})$, for $G$ a Kac--Moody group, in analogy with the classical fundamental domain for $SL_{2}(\mathbb{Z})$ on the Poincar\'e upper half-plane. 

 Finding an exact fundamental domain in the general finite dimensional  case is technically challenging, and has only been worked out explicitly in certain cases. For example, for higher rank groups, Minkowski gave an exact  fundamental domain for $GL_n(\mathbb{Z})$ on $O_n(\mathbb{R})\backslash GL_n(\mathbb{R})$ viewed as the space of positive definite symmetric real quadratic forms ([Mi]). Grenier has given an alternate geometric construction of this fundamental domain ([GGT], [Gr1], [Gr2]). For $n=3$, Grenier's fundamental domain for $GL_3(\mathbb{Z})$  is 5 dimensional and has 1 cusp.

 For our applications, it will suffice to consider an approximate fundamental domain, that is, a fundamental domain constructed using Siegel sets. Such a fundamental domain is generally not exact, but is known to contain the exact fundamental domain.

 For example, the classical Siegel fundamental domain for $SL_2(\mathbb{Z})$ on $SL_2(\mathbb{R})/SO(2)$ is the following `rectangle' in upper-half plane coordinates
$$\{x+iy \mid -1/2 \leq x \leq 1/2, y \geq \sqrt{3}/2\},$$
so it is not exact at the `bottom' but it contains the exact fundamental domain. The bottom boundary circle of the exact fundamental domain is inside this rectangle, not on its boundary.

 An exact fundamental domain for $G(\mathbb{Z})$, when $G$ is a Kac--Moody group appears to be out of reach. However, a fundamental domain in terms of Siegel sets may be obtained.

 In [CGP] the authors prove the following.

\begin{theorem} Let $G$ be a symmetrizable Kac--Moody group of non--affine type.   Let $G'=KA'N$, where $A'$ is the Tits cone. The fundamental domain for $G(\mathbb{Z})$ on $K(G(\mathbb{R}))\backslash G'(\mathbb{R})$ is contained in a single Siegel set that has infinitely many sides indexed over the positive roots of $G$, both real and imaginary. The sides corresponding to imaginary roots  appear with multiplicity according to the dimensions of imaginary root spaces. 
\end{theorem}

 This construction is not a fundamental region over the whole space $K(G(\mathbb{R}))\backslash G(\mathbb{R})$ but rather a subspace, where if $G(\mathbb{R})=KA^+N$ is decomposed in terms of the Iwasawa decomposition, the component corresponding to the Cartan subalgebra is replaced by the Tits cone to obtain $G'=KA'N$. This is also the appropriate region of convergence of Eisenstein series, as we discuss in Subsection~\ref{KMEseries}.

 An analogous result for finite dimensional groups also follows from [CGP]. Namely, if $G$ is a finite dimensional simple algebraic group and $G(\Z)$ is its integer form, then the fundamental domain for $G(\mathbb{Z})$ on $K(G(\mathbb{R}))\backslash G(\mathbb{R})$ is contained in a single Siegel set whose finitely many sides are indexed over the positive roots of $G$.

  The theorem of [CGP]  is obtained in  analogy with Garland's fundamental domain for $G(\Z)$  in the affine case ([G1]). In that case, one gets an additional nicer result that under a natural assumption, the Siegel set breaks into a union of finite pieces indexed over the set of all maximal parabolic subgroups.
 
The restriction of the fundamental domain for $G(\Z)$ to the Cartan subalgebra
 has been studied in [FKH].

\section{Eisenstein series on $SL_2$}\label{Eseries}

 Let $\mathfrak{sl}_2(\mathbb{R})$ denote the Lie algebra of 2$\times$2 matrices of trace 0 over $\mathbb{R}$. Let $SL_2(\mathbb{R})$ denote the group of 2$\times$2 matrices of determinant 1 over $\mathbb{R}$.

  We recall that $SL_2(\mathbb{R})$ acts on the Poincar\'e upper half plane $\mathbb{H}=\{z\in\mathbb{C}\mid Im(z)>0\}$ by fractional linear transformations
$$\left(\begin{matrix}  a & c \\ b & d  \end{matrix}\right)\cdot z\mapsto \dfrac{c+dz}{a+bz},\quad z\in \mathbb{H},\  \left(\begin{matrix}  a & c \\ b & d  \end{matrix}\right)\in SL_2(\mathbb{R}).$$
This defines a right action of $SL_2(\mathbb{R})$ on the Poincar\'e upper half plane
$$z\mapsto z\cdot g$$
satisfying $$g(h\cdot z)=(hg)\cdot z.$$

The following theorem is well known.

\begin{theorem} We have $SL_2(\mathbb{R})=KA^+N$. That is, every $g\in SL_2(\mathbb{R})$ has a unique representation as $g=kan$, $k\in K$, $a\in A^+$, $n\in N$, where
$$N=\left\{\left(\begin{matrix} 1 & x\\ 0 & 1\end{matrix}\right)\mid x\in \R\right\}.$$
$$A^+=\left\{\left(\begin{matrix} r & 0\\ 0& r^{-1} \end{matrix}\right)\mid r\in \R_{>0}\right\},$$
$$ K={SO}_2(\mathbb{R}) = \left\{ \left(\begin{matrix}\cos\theta&-\sin\theta\\ \sin\theta&\cos\theta\\ \end{matrix}\right)\mid 0\leq \theta<2\pi\right\}.$$

\end{theorem}
 The action of  $SL_2(\mathbb{R})$ on $\mathbb{H}$  is transitive and ${SO}_2(\mathbb{R})$ is the stabilizer of the point $i\in \mathbb{H}$. The space of cosets $SL_2(\mathbb{R})/SO_2(\mathbb{R})$ is therefore homeomorphic to $\mathbb{H}$ via the map $$\alpha SO_2(\mathbb{R})\mapsto \alpha i.$$ The coset $1\cdot  SO_2(\mathbb{R})$ corresponds to the point $i$.

 We may also consider the action of the subgroup $\Gamma=SL_2(\mathbb{Z})$ of $SL_2(\mathbb{R})$. The following elements $T$ and $S$ generate $SL_2(\mathbb{Z})$:
$$T=\left(\begin{matrix} 1 & 1\\ 0 & 1\end{matrix}\right):z\mapsto z+1,$$
$$S=\left(\begin{matrix} 0 & 1\\ -1 & 0\end{matrix}\right):z\mapsto -1/z.$$

\subsection{Classical Eisenstein series on the upper half plane}\label{selberg}
 The classical Eisenstein series $E(z, s)$ for $z = x + iy$ in the upper half-plane is given by
$$E(z,s) ={\frac{1}{2}}\sum_{\substack{ (a,b) \in\Z\oplus\Z \\ gcd(a,b)=1}}{\ \frac{Im(z)^s}{|a+bz|^{2s}}}.$$
The series $E(z,s)$ converges absolutely on compact sets for $Re(s) > 1$ and by analytic continuation for other values of $s\in\C$.

 In order to obtain a description of Eisenstein series  which can be generalized to Kac--Moody groups, we first rewrite $E(z, s)$ in terms of the action of 
$\Gamma=SL_2(\mathbb{Z})$ on $\mathbb{H}$. The group $SL_2(\mathbb{R})$ acts on $\mathbb{H}$ by fractional linear transformations
$$\gamma\cdot z\mapsto \dfrac{c+dz}{a+bz}$$
for $\gamma=\left(\begin{matrix} a & c\\ b & d\end{matrix}\right)\in SL_2(\mathbb{R})$ and  $z\in \mathbb{H}$. Then a simple computation shows that
$$Im(\gamma\cdot z)={\frac{Im(z)}{|a+bz|^{2}}}.$$

\begin{proposition} We have
$$E(z,s) ={\frac{1}{2}}\sum_{\gamma\in \Gamma/\Gamma\cap B} Im(\gamma\cdot z)^{s},$$
where $B=\{\left(\begin{matrix} r & b\\ 0 & r^{-1} \end{matrix}\right)\mid r\in\R^{\times}_{>0}, b\in\R\}$. 
\end{proposition}

 {\it Proof:} We have a correspondence
\begin{align*}\Gamma/\Gamma\cap B\quad &\leftrightarrow\quad   \begin{matrix} (a,b) \in\Z\oplus\Z-\{(0,0)\} \\ gcd(a,b)=1\end{matrix}\\
\left(\begin{matrix} a & \ast \\ b & \ast \end{matrix}\right) \quad &\leftrightarrow\quad  (a,b)\end{align*}
since $\Z$ is a principal ideal domain. Hence if $gcd(a,b)=1$, then there exist coprime $(c,d)\in\Z\oplus\Z$ solving the linear Diophantine equation $ad-bc=1$. 

 Next we claim that given a coprime pair $a$ and $b \in \Z$, the pair of integers $c,d$ such that $ad-bc=1$ is unique modulo $B$. Let $c,d\in\Z$ be any pair of integers such that $ad-bc=1$ with $gcd(a,b)=1$. Let $c_0,d_0\in\Z$ be any other such pair. Then
$$ad-bc=1=ad_0-bc_0,$$
so
$$a(d-d_0)-b(c-c_0)=0,$$
that is,
$$a(d-d_0)=b(c-c_0).$$
Since this is an equation involving only integers with $gcd(a,b)=1$, this forces $c-c_0=na$ and $d-d_0=nb$ for some $n\in\Z-\{0\}$. Thus
$$\left(\begin{matrix} a & c \\ b & d\end{matrix}\right) =\left(\begin{matrix} a & c_0+na\\ b & d_0+nb\end{matrix}\right) = \left(\begin{matrix} a & c_0\\ b & d_0 \end{matrix}\right) \left(\begin{matrix} 1 & n\\ 0 & 1\end{matrix}\right)$$
and thus $\left(\begin{matrix} a & c \\ b & d\end{matrix}\right)$ and $ \left(\begin{matrix} a & c_0 \\ b & d_0 \end{matrix}\right)$ differ only by an element of $B$. $\square$

 Our next step is to define a function $\Psi_s$ on $a \in A$ by 
$$\Psi_s(a)=r^{-2s}$$
where $a=\left(\begin{matrix} r& 0\\ 0& r^{-1} \end{matrix}\right)$ and then extend $\Psi_s$ to all of $G$ as follows:
$$\Psi_s(g)=\Psi_s(kan)=\Psi_s(a)=r^{-2s},$$
which is well defined by uniqueness of the Iwasawa decomposition.

 We now use the fact that if $g\in SL_2(\R)$ is in the coset corresponding to a point $z$ in  $\mathbb{H}$, then for any element $\gamma$ of $\Gamma=SL_2(\mathbb{Z})$, we have 
$Im(\gamma\cdot z)^{s}=\Psi_s(\gamma g)$. We obtain thus
$$E(z,s) ={\frac{1}{2}}\sum_{\gamma\in \Gamma/\Gamma\cap B} \Psi_s(\gamma g).$$
We shall see later on that this definition of Eisenstein series on the group $SL_2(\R)$  can be generalized to Kac--Moody groups $G(\R)$.

  We may also use the global $SL_2$--symmetry in type IIB supergravity in $D=10$ spacetime dimensions to describe the coset construction of Eisenstein series. The scalar sector of this theory has  2 real scalar fields, the dilation $\phi$ and the axion $\chi$. These can be combined to define a complex scalar field
$$z=\chi+ie^{-\phi}$$
which parametrizes $SL_2(\mathbb{R})/SO_2(\mathbb{R})\cong \mathbb{H}$ and hence transforms under $SL_2(\mathbb{R})$ as the fractional linear transformation
$$\left(\begin{matrix} a & c \\ b & d\end{matrix}\right):z\mapsto\dfrac{c+dz}{a+bz}.$$
We choose the following coset representative:
$$g(x)=Exp\left[\dfrac{\phi(x)}{2}\left(\begin{matrix} 1 & ~0\\ 0 & -1\end{matrix}\right)\right]Exp\left[\chi(x)\left(\begin{matrix} 0 & 1\\ 0 & 0\end{matrix}\right)\right]=\left(\begin{matrix} e^{\frac{\phi(x)}{2}} & \chi(x)e^{\frac{\phi(x)}{2}}\\ 0 & e^{-\frac{\phi(x)}{2}}\end{matrix}\right)$$
where $h=\left(\begin{matrix} 1 & ~0\\ 0 & -1\end{matrix}\right)$ and $e=\left(\begin{matrix} 0 & 1\\ 0 & 0\end{matrix}\right)$ are the $\mathfrak{sl}_2(\mathbb{R})$ generators and $x$ is a point in spacetime.  This coset representative transforms as
$$g \mapsto k g \gamma.$$
We let 
$$g'=\left(\begin{matrix} e^{\frac{\phi'}{2}} & \chi'e^{\frac{\phi'}{2}}\\ 0 & e^{-\frac{\phi'}{2}}\end{matrix}\right)=
\left(\begin{matrix} e^{\frac{\phi'}{2}} & 0\\ 0 & e^{-\frac{\phi'}{2}}\end{matrix}\right)\left(\begin{matrix} 1 & \chi'
\\ 0 & 1\end{matrix}\right)$$
denote the image of $g$ under $SL_2(\Z)$. Then
$$\Psi_{s}(g')=(e^{\frac{\phi'}{2}})^{-2s}=e^{-\phi's}=[Im(\chi'+ie^{-\phi'})]^{s}.$$

\subsection{Eisenstein series on $SL_2$ using representation theory}

 The seminal example of quantum corrections in type IIB supergravity is given by a function on $SL_2(\Z)\backslash SL_2(\R)/SO(2)$, which is written as a sum over the integer lattice $\Z\oplus\Z-\{(0,0)\}$
$$
\mathcal{E}^{SL_2(\Z)}_s(\mathcal{Z}) = \sum_{\vec{\omega}} \left[ (g \vec{\omega})^\dagger \cdot g \vec{\omega} \right]^{-s}.
$$
The coset representative $g$ of $SL_2(\R)/SO(2)$ is defined in Subsection~\ref{selberg}. Here $\vec{\omega}$ is the fundamental representation of $SL_2(\R)$ restricted to integer values, which defines the integer lattice $\Z\oplus\Z$ that is left invariant under $SL_2(\Z)$ transformations. Setting $\vec{\omega} = m\vec{e}_1 + n\vec{e}_2 = (m,n) \in \Z\oplus\Z$, we have
\begin{align*}
\mathcal{E}^{SL_2(\Z)}_s(\phi,\chi) & = \sum_{(m,n)\in \Z\oplus\Z} \left[ e^\phi \{ (m+n\chi)^2 + n^2 e^{-2\phi} \} \right]^{-s} = \sum_{ (m,n) \in\Z\oplus\Z} { \frac{Im(z)^s}{|m+nz|^{2s}}} \\
& = 2\sum_{k=1}^{\infty} \frac{1}{k^{2s}} \sum_{\substack{(a,b) \in\Z\oplus\Z \\ gcd(a,b)=1}} {\frac{Im(z)^s}{|a+bz|^{2s}}} \\
& = 4 \zeta(2s) E(z,s),
\end{align*}
where $\zeta(2s)$ is the Riemann zeta function and $E(z,s)$ is the Eisenstein series introduced in Section~\ref{selberg}. Note that $(m,n)=k(a,b)$ with $k=gcd(m,n) \in \Z$.

\section{Langlands' construction of Eisenstein series on higher rank groups }

\subsection{Definition of Eisenstein series}

In this subsection, we refer to the classical works of Langlands ([L1]--[L4]). Let $G=G(\R)$ denote a semisimple algebraic group  over $\R$. Let $K$ denote the subgroup invariant under the Chevalley involution. Let $G(\Z)$ denote the $\Z$-form of $G(\R)$. Let $B$ denote the Borel, or upper triangular,  subgroup of $G$. Let $A^+$ denote the diagonal subgroup of $G$ with positive diagonal entries and let $N$ denote the unipotent subgroup of $B$, that is, upper triangular with 1's on the diagonal. Let $G=KA^+N$ denote the Iwasawa decomposition of $G$.

 We have the following result which describes the structure of $A$.
\begin{theorem} ([MP], p 496) Let $A\leq G$ be the subgroup generated by the elements $h_{\alpha_i}(t)$, $t\in \R^{\times}_{>0}$, $i\in I$.  Then every element $a$ of $A$ may be expressed uniquely in the form $a=\prod_{i\in I} h_{\alpha_i}(n_i)$, for $n_i\in \R^{\times}$.
\end{theorem}

  We write 
$$a=\prod_{i=1}^{\ell} h_{\alpha_i}(e^{t_i})$$
where 
$$h_{\alpha_i}(e^{t_i})=e^{t_i\underbar{h}_i}.$$
and $\underbar{h}_i$ are the simple coroots corresponding to $\alpha_i$. 

 Let $g=kan\in G(\R)$, written in Iwasawa form. Let  $s_1,\dots s_{\ell}\in {\mathbb C}^{\times}$. We define 
$$\Psi_{(s_1,\dots ,s_{\ell})} (a)=\prod_{i=1}^{\ell} e^{-2t_is_i},$$
where $e^{t_i}$ are independent characters of the group $G$. Extend $\Psi_{(s_1,\dots ,s_{\ell})}$ to $G$:
$$\Psi_{(s_1,\dots ,s_{\ell})} (g)=\Psi_{(s_1,\dots ,s_{\ell})} (a),$$
and this is well defined since the Iwasawa decomposition is unique. We write $s=(s_1,\dots ,s_{\ell})\in ({\mathbb C}^{\times})^{\ell}$ and $r_i=e^{t_i}$.

 Langlands extended Selberg's definition of Eisenstein series to higher dimensional groups as follows  ([L1]--[L4]). For $g\in K(G) \backslash G(\R)$ and $\gamma \in G$, define the minimal parabolic Eisenstein series on the double coset space $K(G) \backslash G(\R)/G(\Z)$ in the following way
$$
\eqnlab{EisensteinDef}
E_{s}(g)=\frac{1}{2} \sum_{\gamma\in( \Gamma/\Gamma\cap{B})}\quad 
\Psi_{s}(g\gamma).
$$

 This is defined in analogy with Eisenstein series on $SL_2(\R)$. We can also define a maximal parabolic  Eisenstein series by replacing $B$ with a maximal parabolic subgroup $P_i$.

Then  $E_{s}(g)$ converges absolutely in the region $Re(\Psi_{s}(h_{\alpha_i}(r_i)))< -2$ for each $i$. This is analogous to the classical condition $Re (s_i)>1$.

 The number of linearly independent characters is equal to the number of simple roots which also equals  the number of fundamental representations. The order numbers $s = (s_1,\dots,s_{\ell})$ thus label the simple roots and hence can be given an interpretation in terms of Dynkin labels. By `Dynkin labels',  we mean a set of $\ell$ integers assigned to the fundamental weights.


 We have discussed the most general form of a character. Those characters appearing so far in string theory seem to be of a special form corresponding to special values of $s$. However, one cannot rule out all the Eisenstein series that correspond to other values of $s$, since only certain special terms in supergravity have been investigated (as in [GMRV]). {\footnote
{The definition used in [GMRV]
$$
E_\lambda(g) = {\frac{1}{2}}\sum_{\gamma\in \Gamma/\Gamma\cap B} e^{(\lambda+\rho)(H(g \gamma))}
$$
is an alternative notation for Eisenstein series where $s = \lambda + \rho$. More precisely,
$
e^{s_i(H(g \gamma))} = h_{\alpha_i} (r_i(g \gamma))^{-2s_i}.
$
The shift by the Weyl vector $\rho$ in the definition simplifies the expression of the constant term of the Eisenstein series.}}

\subsection{Example: $G=SL_n(\R)$}  Let $g\in G$. Then $g$ has a unique Iwasawa decomposition $g = kan$. For $G=SL_2(\mathbb{R})$, we defined the character as
$$
\Psi_s (g) = \Psi_s (a) = r^{-2s}
$$
where $a = \left(\begin{matrix}  r & 0 \\ 0 & r^{-1}  \end{matrix}\right)$. It is straightforward to generalize this definition to $SL_n(\mathbb{R})$, in which case the diagonal component can be decomposed according to
$$
a = \left(\begin{matrix}  r_1 & 0 & 0 & \cdots & 0 \\ 0 & r_1^{-1}r_2 & 0 & \cdots & 0 \\ 0 & 0 & r_2^{-1}r_3 & \cdots & 0 \\ \vdots & \vdots & \vdots & \ddots & \vdots \\ 0 & 0 & 0 & \cdots & r_{n-1}^{-1}  \end{matrix}\right) = \left(\begin{matrix}  r_1 & 0 & 0 & \cdots & 0 \\ 0 & r_1^{-1} & 0 & \cdots & 0 \\ 0 & 0 & 1 & \cdots & 0 \\ \vdots & \vdots & \vdots & \ddots & \vdots \\ 0 & 0 & 0 & \cdots & 1  \end{matrix}\right) \left(\begin{matrix}  1 & 0 & 0 & \cdots & 0 \\ 0 & r_2 & 0 & \cdots & 0 \\ 0 & 0 & r_2^{-1} & \cdots & 0 \\ \vdots & \vdots & \vdots & \ddots & \vdots \\ 0 & 0 & 0 & \cdots & 1  \end{matrix}\right) \dots .
$$
There are now $(n-1)$ independent characters, and one can define
$$
\Psi_s (g) = \Psi_s (a) = r_1^{-2s_1} \dots r_{n-1}^{-2s_{n-1}} = \prod_{i = 1}^{n-1} r_i^{-2s_i}
$$
where $s = (s_1, \dots, s_{n-1})$ is understood as a vector of order numbers $s_i$. Since $r_i$ is the contribution from the Cartan generator $h_{\alpha_i}$corresponding to the simple root $\alpha_i$, we can also write the above expression as
$$
\Psi_s (a) = \prod_{i = 1}^{n-1} h_{\alpha_i} (r_i)^{-2s_i}.
$$
The number of characters is equal to the rank of the group $G(\mathbb{R})$, and is independent of the dimension of the representation.

\subsection{Minimal co--adjoint orbits}

 Let $\Mg$ be a finite dimensional simple Lie algebra. The adjoint representation of $\Mg$ can be uniquely decomposed in terms of its $\mathfrak{sl}_2 \times \mathfrak{u}$ subalgebra, with $\mathfrak{u}$ being the largest subalgebra of $\Mg$ commuting with $\mathfrak{sl}_2$, according to:

$$\Mg = 1 \oplus R  \oplus  \{ 1 \oplus Adj(\mathfrak{u}) \} \oplus  R  \oplus  1 = \Mg_{-2}  \oplus  \Mg_{-1}  \oplus  \Mg_0  \oplus  \Mg_1  \oplus  \Mg_2$$

 where $R$ is a  representation of $\mathfrak{u}$. We may also label the generators as follows:

$$\Mg = \mathfrak{f}  \oplus  \Mg_{-1}  \oplus  (\mathfrak{h}  \oplus  Adj(\mathfrak{u}))  \oplus  \Mg_1  \oplus  \mathfrak{e}.$$

 In particular, $\{\mathfrak{e}, \mathfrak{h}, \mathfrak{f} \}$ correspond to the above mentioned $\mathfrak{sl}_2$ subalgebra spanned by the highest and lowest root.

The dimension of the minimal co--adjoint orbit is known to be ([J])

$$d = \frac{1}{2}( dim(\Mg) - dim(\mathfrak{u}) + rank(\Mg) - rank(\mathfrak{u}) ).$$

 Moreover, the minimal co--adjoint orbit can be obtained by computing the  action on the highest weight vector of the adjoint representation, that is, the action on the highest root ([KPW]).

 In fact, the co--adjoint\footnote{Using the Killing form, the adjoint and the co--adjoint orbits are isomorphic.} orbit of the highest root is given by the direct sum of spaces
 $$\C \mathfrak{h} \oplus  G_1 \oplus \C \mathfrak{e}$$
where  $G_1$ is $\C$--vector space with the positive roots as basis. By [KPW], this is the minimal orbit. The stabilizer of this orbit is generated by
$$\{Adj(\mathfrak{u}), \Mg_{-1}, \Mg_{-2}\}.$$ 
and coincides with the Lie algebra of $P_{\alpha_1}$ for simply laced simple Lie algebras ([KPW], [GMV]).

 For $\Mg=\mathfrak{e}_{6(6)}$,  $\mathfrak{u} = \mathfrak{sl}_6$ and we decompose $\mathfrak{e}_{6(6)}$ as
$${\bf 78} = {\bf 1}\oplus {\bf 20} \oplus\{ {\bf 35}\oplus {\bf 1} \}\oplus {\bf 20}\oplus {\bf 1}.$$

 For $\Mg=\mathfrak{e}_{8(8)}$, $\mathfrak{u} = \mathfrak{e}_{7(7)}$. The
adjoint representation of $\mathfrak{e}_{8(8)}$ decomposes as 
$${\bf 248} = {\bf 1}\oplus   {\bf 56} \oplus  \{{\bf 133} \oplus  {\bf 1}\} \oplus  {\bf 56} \oplus  {\bf 1}$$
which corresponds to
$$\Mg = \Mg_{-2}  \oplus  \Mg_{-1}  \oplus  \Mg_0  \oplus  \Mg_1  \oplus  \Mg_2.$$

\subsection{Remarks on Langlands' construction}

Langlands' construction of a maximal parabolic Eisenstein series on $G(\Z)$ is given by:  ([L1]--[L4])
$$
E_{P_{\alpha_i},s}(g) = \frac{1}{2} \sum_{\gamma \in (\Gamma/\Gamma \cap P_{\alpha_i})} \Psi_s(g\gamma).
$$

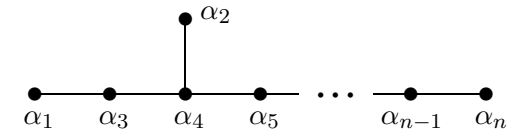
\begin{figure}[ht!]
\begin{center}
\setlength{\unitlength}{1mm}
\begin{picture}(60,10)
   \put(0,0){\circle*{1.7}}
   \put(20,10){\circle*{1.7}}
   \put(10,0){\circle*{1.7}}
   \put(20,0){\circle*{1.7}}
   \put(30,0){\circle*{1.7}}
   \put(38,-0.1){\circle*{0.6}}
   \put(40,-0.1){\circle*{0.6}}
   \put(42,-0.1){\circle*{0.6}}
   \put(50,0){\circle*{1.7}}
   \put(60,0){\circle*{1.7}}

   \put(0,0){\line(1,0){35}}
   \put(45,0){\line(1,0){15}}
   \put(20,0){\line(0,1){10}}

   \put(-1.5,-4){$\alpha_1$}
   \put(22,10){$\alpha_2$}
   \put(8.5,-4){$\alpha_3$}
   \put(18.5,-4){$\alpha_4$}
   \put(28.5,-4){$\alpha_5$}
   \put(46,-4){$\alpha_{n-1}$}
   \put(58.5,-4){$\alpha_n$}
\end{picture}
\end{center}
  \caption{The Dynkin diagram of $E_{n}$.}
  \label{fig:DynkinEn}
\end{figure}

For $\alpha_i$ a simple root, the parameter $s$ takes the value $s=2\hat{s}\omega_i$ where $\omega_i$ is the fundamental weight corresponding to $\alpha_i$, and $\hat{s}$ is an arbitrary complex parameter ([GMRV]).

The information in Table 6 of [GMV] contains some of the ingredients that may allow a generalization the notion of minimal co--adjoint orbit to Kac--Moody groups.


The highest root vectors for simply laced semisimple Lie algebras are given in Table~\ref{tab:adjoint}. The weight vectors of the adjoint representation,  in most cases, do not lie  in the fundamental representation $\omega_1 = (1 0 \dots 0)$. Moreover, the dimension of the adjoint representation is often larger than that of the fundamental representation $\omega_1$. Various representations of the groups of interest here are summarized in Table~\ref{tab:repr}. 

\begin{table}[ht]
\centering
\begin{tabular}{|c|l|l|}
\hline
$G$ & Coxeter labels & Dynkin labels \\
\hline
\hline
$A_1$ & 1 & 2 \\
$A_4$ & $[1 1 1 1]$ & $(1 1 0 0)$ \\
$D_5$ & $[1 1 2 2 1]$ & $(0 0 1 0 0)$ \\
$E_6$ & $[1 2 2 3 2 1]$ & $(0 1 0 0 0 0)$ \\
$E_7$ & $[2 2 3 4 3 2 1]$ & $(1 0 0 0 0 0 0)$ \\
$E_8$ & $[2 3 4 6 5 4 3 2]$ & $(0 0 0 0 0 0 0 1)$ \\
\hline
\end{tabular}
\caption{Highest weight vectors of the adjoint representation of some Lie groups.}
\label{tab:adjoint}
\end{table}

\begin{table}[ht]
\centering
\begin{tabular}{|c|l|l|l|}
\hline
$G$ & Basic representations & Defining representation & Adjoint representation \\
\hline
\hline
$A_1$ & $1$ & $1$ & $2$ \\
$A_1 \times A_1$ & $(1;1)$ & $(1;1)$ & $(2;2)$ \\
$A_2 \times A_1$ & $(1 0;1)$ & $(1 0;1)$ & $(1 1;2)$ \\
$A_4$ & $(1 0 0 0)$ & $(1 0 0 0)$ & $(1 1 0 0)$ \\
$D_5$ & $(0 1 0 0 0), (0 0 0 0 1)$ & $(1 0 0 0 0)$ & $(0 0 1 0 0)$ \\
$E_6$ & $(1 0 0 0 0 0)$ & $(1 0 0 0 0 0)$ & $(0 1 0 0 0 0)$ \\
$E_7$ & $(0 0 0 0 0 0 1)$ & $(0 0 0 0 0 0 1)$ & $(1 0 0 0 0 0 0)$ \\
$E_8$ & $(0 0 0 0 0 0 0 1)$ & $(0 0 0 0 0 0 0 1)$ & $(0 0 0 0 0 0 0 1)$ \\
\hline
\end{tabular}
\caption{Highest weight representations of some Lie groups in Dynkin labels.}
\label{tab:repr}
\end{table}

\section{Construction of Eisenstein series using representation theory}\label{EseriesRep}

Our strategy here is to start with the fundamental representation of a simple Lie algebra, and explore the role of  tensor products of the representation with itself in the definition of Eisenstein series. The formulas are known explicitly for the basic representations of $\mathfrak{sl}_n$, and should generalize to the basic representations of all simply laced simple Lie algebras. The corresponding results for Kac--Moody algebras are not yet known.

Assume for now that the Lie algebra $\mathfrak{g}$ has only one `basic module', which generates all others, denoted $V^{\lambda_1}$. The underlying vector space of this module is spanned by a set of orthonormal basis $\{v_i\}$, such that $v_1=(1, 0, \dots, 0)$ represents the highest weight vector.

\subsection{Eisenstein series on $SL_2$ revisited}

Before giving the general formulas, we will first work out the details for $\mathfrak{sl}_2$. The basic representation of $\mathfrak{sl}_2(\R)$ is the two dimensional fundamental representation denoted $\mathbf{2}$. Taking the tensor product of this representation with itself we obtain $\mathbf{2} \times \mathbf{2} = \mathbf{3} \oplus \mathbf{1}$, where $\mathbf{3}$ is the adjoint representation and $\mathbf{1}$ is the trivial representation. 

Let $\gamma^{(2\times 2)}$ denote the tensor product of the image of the group element $\gamma \in \Gamma$ with itself in the basic representation
$$\gamma^{(2\times 2)} = \left( \begin{array}{cc}
e^{\phi/2} & \chi e^{\phi/2} \\
0 & e^{-\phi/2} \\
\end{array} \right) \otimes \left( \begin{array}{cc}
e^{\phi/2} & \chi e^{\phi/2} \\
0 & e^{-\phi/2} \\
\end{array} \right).$$
Let $g^{(2\times 2)}$ denote the tensor product of the image of the coset representative $g \in SO(2)\backslash SL_2(\R)$ with itself in the basic representation
$$g^{(2\times 2)} = \left( \begin{array}{cc}
a & c \\
b & d \\
\end{array} \right) \otimes \left( \begin{array}{cc}
a & c \\
b & d \\
\end{array} \right).$$
Using the transformation matrix
$$M=\left( \begin{array}{cccc}
1 & 0 & 0 & 0 \\
0 & \frac{1}{\sqrt{2}} & 0 & \frac{1}{\sqrt{2}} \\
0 & \frac{1}{\sqrt{2}} & 0 & -\frac{1}{\sqrt{2}} \\
0 & 0 & 1 & 0 \\
\end{array} \right),$$
both $M^{-1}\gamma^{(2\times 2)}M$  and  $M^{-1}g^{(2\times 2)}M$ are block--diagonal so that the first $3 \times 3$ block corresponds to the adjoint representation and the second $1 \times 1$ block is the trivial representation.

The highest weight vector of the adjoint module is given by $v_1^{(2\times 2), \text{adj}}=(1, 0, 0, 0)$. The summand of the Eisenstein series can then be shown to satisfy
$$
\Psi_{s} (g \gamma) = ||g \gamma \cdot v_1||^{-2s} = ||g \gamma \cdot v_1^{(2\times 2), \text{adj}}||^{-s}.
$$
In fact, in the above summand we could have dropped the $1 \times 1$ block corresponding to the trivial representation in $g^{(2\times 2)}$ and $\gamma^{(2\times 2)}$ due to their block-diagonal form to obtain
$$
\Psi_{s} (g \gamma) = ||g \gamma \cdot v_1||^{-2s} = ||g \gamma \cdot v_1^{\text{adj}}||^{-s}.
$$

Let us continue by looking at the tensor product $\mathbf{3} \times \mathbf{3} = \mathbf{5} \oplus \mathbf{3} \oplus \mathbf{1}$. The procedure is the same as before, but now with the resulting product being the direct product of three representations. This means that  by using the transformation matrix 
$$M=\left( \begin{array}{ccccccccc}
1 & 0 & 0 & 0 & 0 & 0 & 0 & 0 & 0 \\
0 & \frac{1}{\sqrt{2}} & 0 & 0 & 0 & \frac{1}{\sqrt{2}} & 0 & 0 & 0 \\
0 & 0 & \frac{1}{\sqrt{6}} & 0 & 0 & 0 & \frac{1}{\sqrt{2}} & 0 & \frac{1}{\sqrt{3}} \\
0 & \frac{1}{\sqrt{2}} & 0 & 0 & 0 & -\frac{1}{\sqrt{2}} & 0 & 0 & 0 \\
0 & 0 & \sqrt{\frac{2}{3}} & 0 & 0 & 0 & 0 & 0 & -\frac{1}{\sqrt{3}} \\
0 & 0 & 0 & \frac{1}{\sqrt{2}} & 0 & 0 & 0 & \frac{1}{\sqrt{2}} & 0 \\
0 & 0 & \frac{1}{\sqrt{6}} & 0 & 0 & 0 & -\frac{1}{\sqrt{2}} & 0 & \frac{1}{\sqrt{3}} \\
0 & 0 & 0 & \frac{1}{\sqrt{2}} & 0 & 0 & 0 & -\frac{1}{\sqrt{2}} & 0 \\
0 & 0 & 0 & 0 & 1 & 0 & 0 & 0 & 0 \\
\end{array} \right),$$
the matrices $M^{-1}g^{(3\times 3)}M$ and $M^{-1}\gamma^{(3\times 3)}M$ are simultaneously block--diagonal with three blocks each. We have now two choices for the highest weight vector: 
$$v_1^{(3\times 3), 5}=(1, \underbrace{0, \dots, 0}_{8})$$ or 
$$v_1^{(3\times 3), \text{adj}}=(\underbrace{0, \dots, 0}_{5}, 1, 0, 0, 0).$$  It can be shown that
$$
\Psi_{s} (g \gamma) = ||g \gamma \cdot v_1^{(3\times 3), 5}||^{-\frac{s}{2}} = ||g \gamma \cdot v_1^{5}||^{-\frac{s}{2}} = ||g \gamma \cdot v_1^{(3\times 3), \text{adj}}||^{-s} = ||g \gamma \cdot v_1^{\text{adj}}||^{-s}.
$$
Thus, choosing the highest weight vector of a certain representation ensures that the summand remains inside the same representation. 

Relative to a choice of representation, we obtain an associated `effective' order number in that representation. In the examples above, we have $s^{\text{adj}}=\frac{s}{2}$ and $s^{5}=\frac{s}{4}$ where $s$ is the order number in the basic representation. The order number $s$ can thus be given a representation theoretic interpretation. In other words, we may view $\Psi_{s}$ from the perspective of the basic representation with order number $s$, or equivalently view it as a different representation an associated effective order number.



\subsection{Eisenstein series from fundamental representations}\label{fund}

 We now generalize the construction of Eisenstein series using representation theory to all simply laced semisimple algebraic groups and Kac--Moody groups. We start the construction by choosing a highest weight module $V$. When the Lie algebra $\Mg$ is finite dimensional, $V$ can be taken to be the adjoint representation. However, a common choice for $V$ is the $\omega_1$--fundamental representation ([LW2], [CG2]). When the root lattice equals the weight lattice, the $\omega_1$--fundamental representation and the $\alpha_1$--fundamental representation coincide. Note that when $\Mg$ is a Kac--Moody algebra, the adjoint representation is not a highest weight representation. We may also choose $V$ to be a tensor product of fundamental representations as needed. 

 Beyond the modular group, the construction of Eisenstein series using a $G(\Z)$--invariant lattice was first derived for $SL_3(\Z)$ in [KiP] and [GV]. In [KiP], the $SL_5(\Z)$ case was also discussed. Later, the construction for general $SL_n(\Z)$ and $SO(d,d;\Z)$ was given in [OP1]. We note that Eisenstein series on $Spin(d,d)$ can be obtained from Eisenstein series on $SO(d,d)$ ([GMV]). 

 Since the parameter space of an automorphic form is a symmetric space, it is possible to construct the non-holomorphic Eisenstein series using a representative of the coset. Let $g$ be a representative of the coset $K(G)\backslash G$ transforming according to
$$
   g \longmapsto k g \gamma, \hspace{1cm} \gamma \in G \quad \text{and} \quad k \in K(G).
$$
Let $\vec{\omega}, \vec{\omega}' \in V_{\Z}$, then
$$
   g \vec{\omega} \longmapsto k g \vec{\omega}'
   \eqnlab{VwTransformation}
$$

 under the $G(\mathbb{Z})$ action. We can thus form the non-holomorphic Eisenstein series as
$$
   E_s(g) = \sum_{\vec{\omega}\in V_{\Z}} \delta(f(\vec{\omega}))) \left[ (g \vec{\omega})^\dagger \cdot g \vec{\omega} \right]^{-s},
   \eqnlab{LatticeAutomorphic}
$$

 which by construction, is invariant under $G(\mathbb{Z})$. The coset transformation rule shows that invariance under $G(\Z)$ requires $k^\dagger k = 1$, which is satisfied since $K(G)$ is the unitary form. This Eisenstein series is non-holomorphic due to the appearance of the Hermitian conjugate.

 The `constraint factor' $\delta(f(\vec{\omega})))$ reduces the sum to a single $G(\Z)$--orbit on $V_{\Z}$. There is only one $G(\Z)$--orbit on the basic modules of $\mathfrak{sl}_n$ and therefore the constraint factor is not needed there. However, representations of $\mathfrak{sl}_n$ which are antisymmetric tensors will likely require constraints. The vector representation of $SO(d,d)$ requires the quadratic constraint $\vec{\omega}^\dagger \wedge \vec{\omega} = \kappa$ with a constant $\kappa$, which picks out a single $G(\Z)$--orbit (see [OP1] and [AFP]). For higher dimensional modules or higher rank groups, more complicated constraints may be needed in order to reduce the sum to a single $G(\Z)$--orbit.

 The relationship between the quadratic constraint and the choice of maximal parabolic subgroup is noted in [GMV]. Langlands' construction of Eisenstein series is defined as a sum over a single $G(\Z)$--orbit  ([L1]--[L4]). Choosing a parabolic subgroup corresponds therefore to choosing a particular orbit. The constraints appearing in the construction using representations theory ensure the sum is taken over $\gamma \in \Gamma/\Gamma \cap B$. The quadratic constraint for $Spin(d,d)$ becomes then quite natural, since the parabolic subgroup stabilizes a null vector.

 With the appropriate constraint, the Eisenstein series defined in this subsection becomes an eigenfunction of all the Casimir operators of $G$. There are a number of open questions regarding the use of constraints. We discuss these briefly in Section~\ref{conclusions}.

 The construction of the Eisenstein series in [OP1] and [LW] can be shown to coincide with Langlands' construction up to constraints that define a particular $G(\Z)$--orbit. Let $V$ be a highest weight module for $\Mg$ which is endowed with a positive Hermitian inner product $\{,\}$. This module is spanned by an orthonormal basis $\{v_i\}$ such that $\{v_i,v_j\}=\delta_{ij}$. Let $G$ be the corresponding Lie group or Kac--Moody group over $\R$. We assume the Iwasawa decomposition of a group element $g \in G$ is such that $g = kan$, where $k \in K(G)$ is unitary with respect to $\{,\}$,  $a \in A^+$ is diagonal with positive entries  and $n \in N$ is upper triangular with 1's on the diagonal. We note that Iwasawa decomposition is well known in the finite dimensional case and was proven for affine Kac--Moody groups in [G5] and general Kac--Moody groups over arbitrary fields in [DGH].

The linear space $V$ can be decomposed into mutually orthogonal subspaces $V = \bigoplus_\mu V_{\mu}$, where $\mu \in \mathfrak{h}^{\ast}$. A Cartan generator $a\in A$ acts  on $V_{\mu}$ as
$$
a \cdot v = a^{\mu} v, \quad a \in A, \, v \in V_{\mu}.
$$
That is, $V_{\mu}$ is the eigenspace of $A$ with eigenvalue $a^{\mu}$. If $V$ is the defining representation, the integer form $V_{\mathbb{Z}} \subseteq V$can be taken to be the space spanned by the basis of $V$ with integer coefficients. Assuming $v_i \in V_{\mathbb{Z}}$
$$
\sum_{i=1}^n a_i v_i \in V_{\mathbb{Z}} \quad \Leftrightarrow \quad a_i \in \mathbb{Z}.
$$
Since $\{v_i\}$ is an orthonormal basis, we have the implication
$$
w, w' \in V_{\mathbb{Z}} \quad \Rightarrow \quad \{w,w'\} \in \mathbb{Z}.
$$
Choose a basis $\{v_i\}$ for $V$ that is coherently ordered relative to depth.  When $V$ is the defining representation, we may write these basis vectors as column vectors which in the finite dimensional case can be represented
$$
v_1 = \left(\begin{matrix}  1 \\ 0 \\ 0 \\ \vdots \\ 0  \end{matrix}\right), \, v_2 = \left(\begin{matrix}  0 \\ 1 \\ 0 \\ \vdots \\ 0  \end{matrix}\right), \, v_i = \left(\begin{matrix}  0 \\ \vdots \\ 1 \\ \vdots \\ 0  \end{matrix}\right) \leftarrow i\text{-th component}, \dots, v_n = \left(\begin{matrix}  0 \\ \vdots \\ 0 \\ \vdots \\ 1  \end{matrix}\right)
$$
and in the Kac--Moody case as 
$$
v_1 = \left(\begin{matrix}  1 \\ 0 \\ 0 \\ \vdots \\   \end{matrix}\right), \, v_2 = \left(\begin{matrix}  0 \\ 1 \\ 0 \\ \vdots \\   \end{matrix}\right), \, v_i = \left(\begin{matrix}  0 \\ \vdots \\ 1 \\ \vdots \\   \end{matrix}\right) \leftarrow i\text{-th component}, \dots .
$$

 In this convention, $v_{\lambda}=v_1$ is  the highest weight vector. The rest of the weights of this representation are of the form
$$
v_1 - \sum_{j=1}^{\ell} \kappa_j \alpha_j, \quad \kappa_j \in \mathbb{Z}_{\ge 0}, \quad \ell=dim_{\C}(\mathfrak{h}),
$$
with $\alpha_j$ being the simple roots embedded in $V$.
 We  define the integer form $\Gamma=G(\Z) \subseteq G(\R)$
$$
\Gamma =G(\Z)= \{ \gamma \in G(\R) \mid \gamma\cdot V_{\mathbb{Z}} = V_{\mathbb{Z}} \}.
$$
Let $\Gamma$ act on $v_1$ by the standard left action $(\gamma \cdot v_1)$. This does not generate all of $V_{\Z}$ but if we take the orbit $\mathcal{U}_{\Z}\cdot v_1$ we obtain all of $V_{\Z}$. 

 For $g=kan$, we have $\Psi_s(g)=||gv_{\lambda}||^{-2s}$ as one of the independent characters. The action of $g$ on $v_{\lambda}$ can be described as follows:

 $n$ stabilizes $v_{\lambda}$,

 $a$ acts on  $v_{\lambda}$ as scalar multiplication by $\langle \lambda,log(a)\rangle\in\mathfrak{h}$,

 $k$ is norm preserving.

 Then $\Psi_s$ is right $N$--invariant and left $K$--invariant.

 For $g\gamma$ with $g \in K(G) \backslash G(\mathbb{R})$ and $\gamma \in \Gamma$ the character corresponding to $s_1$ is
$$
\eqnlab{CharRep}
\Psi_{s_1} (g \gamma) = ||g \gamma \cdot v_1||^{-2s_1} = \{ g\gamma \cdot v_1, g\gamma \cdot v_1 \}^{-s_1}.
$$

 The Eisenstein series can be rewritten as
$$
E_{s_1}(g) = {\frac{1}{2}}\sum_{\gamma\in \Gamma/\Gamma\cap B} \Psi_{s_1}(g\gamma) = {\frac{1}{2}}\sum_{\gamma\in \Gamma/\Gamma\cap B} \{ g(\gamma \cdot v_1), g(\gamma \cdot v_1) \}^{-s_1},
$$
where the expression after the second equality is given in [OP1] up to the orbit selecting constraints.

 Assume that the diagonal part of $g\gamma = kan$ takes the following form
$$
a = \left(\begin{matrix}  a_1 & 0 & 0 & \cdots & 0 \\ 0 & a_2 & 0 & \cdots & 0 \\ 0 & 0 & a_3 & \cdots & 0 \\ \vdots & \vdots & \vdots & \ddots & \vdots \\ 0 & 0 & 0 & \cdots & a_n  \end{matrix}\right).
$$
This construction using representation theory explicitly gives a clean expression for the character $a_1$
$$
a \cdot v_1 = a_1 v_1.
$$

\subsection{Constructing general Eisenstein series using tensor products}

 For $\mathfrak{sl}_n(\mathbb{R})$ all other fundamental representations can be constructed from the defining representation $V$ by  taking exterior products. For instance,
$$
\{v_1 \wedge v_2, v_1 \wedge v_3, \dots\}
$$
is the basis of another fundamental representation, which extracts the character
$$
a \cdot (v_1 \wedge v_2) = a_1 a_2 (v_1 \wedge v_2).
$$
Since $v_1 \wedge \dots \wedge v_n \propto 1$, there are in total $(n-1)$ fundamental representations as expected. The $n$ linearly independent characters obtained by this method are
$$
a_1^{-2s_1}, \, (a_1 a_2)^{-2s_2}, \, (a_1 a_2 a_3)^{-2s_3}, \, \dots.
$$

  For finite dimensional simple Lie algebras, all fundamental representations are obtained from a subset of given ones, called basic modules. In the simply laced case, there is a single basic module.

  For types $C$ and $E$ algebras the basic module is simply the defining module, that is, the fundamental module of smallest dimension. For type $A$ algebras, either of the two defining modules can be taken as a basic module. For type $B$ algebras, the basic module  is the spinor module, which is a fundamental module other than the defining one. For type $D$ algebras, as basic modules we take two distinct fundamental modules (spinor and conjugate spinor), which are both defining modules (see [FS], page 235). 

 However, it is unclear if one may generate all highest weight modules in this way for  Kac--Moody algebras.

 In the following we will assume that all the fundamental modules are known. We name them $V^{\lambda_1}, \dots, V^{\lambda_{\ell}}$, where $\lambda_i$ denotes the highest weight vector of the module $V^{\lambda_i}$ and $\ell$ is the rank of $G$. We can determine one character for each module $V^{\lambda_i}$ using the procedure described in Section~\ref{fund}. The most general Eisenstein series is then defined as
$$
\eqnlab{EGeneralS}
E_{s}(g) = {\frac{1}{2}}\sum_{\gamma\in \Gamma/\Gamma\cap B} \prod_{i=1}^{\ell} ||g \gamma \cdot v_{\lambda_i}||^{-2s_i} = {\frac{1}{2}}\sum_{\gamma\in \Gamma/\Gamma\cap B} \prod_{i=1}^{\ell} e^{-2 s_i t_i(\gamma)}.
$$

 The exponent in this definition is proportional to $\sum_i s_i t_i$ with completely arbitrary order numbers $s_i \in \C$.

 Instead of explicitly computing all characters corresponding to different fundamental modules, we may also construct a highest weight module $V$ as a tensor product of the fundamental modules and then define Eisenstein series relative to this single module. Let $V^{\hat{\lambda}}$ be a highest weight module with  highest weight the dominant integral weight $\hat{\lambda}=\sum_{i=1}^{\ell} m_i \lambda_i$. Then we have
$$
V^{\hat{\lambda}} = V^{(\sum_{i=1}^{\ell} m_i \lambda_i)} = \bigotimes_{i=1}^\ell (V^{\lambda_i})^{\otimes m_i}.
$$
Since
$$
a \cdot v_{\hat{\lambda}} = \prod_{i=1}^{\ell} a_i^{m_i} v_{\hat{\lambda}},
$$
we define the Eisenstein series on the module $V^{\hat{\lambda}}$ as
$$
E_{\hat{s}}(g) = {\frac{1}{2}}\sum_{\gamma\in \Gamma/\Gamma\cap B} ||g \gamma \cdot v_{\hat{\lambda}}||^{-2\hat{s}} = {\frac{1}{2}}\sum_{\gamma\in \Gamma/\Gamma\cap B} \prod_{i=1}^{\ell} e^{-2 m_i \hat{s} t_i(\gamma)}.
$$
This way of defining an Eisenstein series involves only one order parameter $\hat{s}$. However, it is not possible to obtain the most general Eisenstein series using this method, since $s_i \in \C$ while $m_i \in \mathbb{Z}_{\ge 0}$ and $\hat{s} \in \C$. The restriction that $m_i$ are non-negative integers follows from the fact that $\hat{\lambda}$ is a weight vector.

The expression $E_{\hat{s}}(g)$ can be rewritten further in terms of only the basic modules, but this does not appear to give any further insight.

By our construction, $\sum_{\gamma\in\Gamma/\Gamma\cap B }$ is a sum over the $\Gamma$--orbit of a highest weight vector $v_1$, not a sum over all of  $V_{\Z}$ in general. As we have mentioned, constraints may be required on the sum in $E_{\hat{s}}(g)$. In representation theoretic terms, a constraint may be given by a projection onto an irreducible component in the tensor product that defines the lattice $G(\Z)$, as in [OP1]. In particular, the projection onto a single orbit is investigated representation theoretically in [OP1] in terms of the projection of the tensor product $V\otimes V$ onto its largest component.

We have shown that the order numbers $s$ can be given a representation--theoretic interpretation. The choice of representation determines the highest weight vector $v$ in the expression $||g\gamma \cdot v||^{-2s}$, while the minimal co--adjoint orbit is related to the $G(\Z)$--orbit of $v\in V_{\Z}$. It remains to determine the precise relationship between the choice of representation and the minimal co--adjoint orbit. For example, it is not fully understood how the choice of values for the order numbers $s$ affects the orbit structure.

\section{Eisenstein series on Kac--Moody groups}

\subsection{Eisenstein series on affine Kac--Moody groups}\label{aff}

\medskip \noindent Let $\widehat{G}$ be a complete affine Kac--Moody group over $\R$. The following construction is taken from [G2] and [G3]. Let $\lambda$ be a dominant integral weight and let $V^{\lambda}$ be the corresponding irreducible highest weight module. Let $V^{\lambda}_{\Z}$ be a Chevalley lattice in $V^{\lambda}$. Let $\{\cdot,\cdot\}$ be the positive definite Hermitian inner product on  $V^{\lambda}_{\Z}$ as in Subsection~\ref{Zform}.  Then

\medskip \noindent (1) $\{v,w\}\in\Z$ for all $v$, $w\in V^{\lambda}_{\Z}$

\medskip \noindent (2) $||v_{\lambda}||=\{v_{\lambda},v_{\lambda}\}=1$ for a highest weight vector $v_{\lambda}$, where $V^{\lambda}_{\lambda,\Z}=\Z\cdot v_{\lambda}$.

\medskip \noindent (3) $\{V^{\lambda}_{\mu,\Z},V^{\lambda}_{\mu',\Z}\}=0$ for weights $\mu$, $\mu'$ of $V^{\lambda}$, $ \mu \neq \mu'$.

\medskip \noindent  Let $V^{\lambda}_{\R}=\R\otimes_{\Z}V^{\lambda}_{\Z}$. Then $\widehat{G}$ is a subgroup of $Aut(V^{\lambda}_{\R})$. Let $\widehat{K}$ be the unitary subgroup of $\widehat{G}$ relative to $\{\cdot,\cdot\}$. Then $\widehat{K}$ plays the role of the compact  subgroup of $\widehat{G}$, though $\widehat{K}$ is not compact.

\medskip \noindent Let $\Psi=\{v_1,v_2,\dots\}$  be a coherently ordered basis of $V^{\lambda}_{\Z}$. With respect to $\Psi$, we obtain an infinite dimensional matrix representation of $\widehat{G}$. Let $\widehat{B}$ be the  upper triangular subgroup of $\widehat{G}$ relative to $\Psi$, let  $\widehat{N}$ be the unipotent subgroup, that is, upper triangular with 1's on the diagonal, let $\widehat{A}$ be the `positive' diagonal subgroup, with positive diagonal entries.

\medskip \noindent Let $\widehat{\Gamma}$ be an `arithmetic' subgroup of $\widehat{G}$. That is, 
$$\widehat{\Gamma}=\{\gamma\in \widehat{G}\mid \gamma \cdot V^{\lambda}_{\Z}= V^{\lambda}_{\Z}\}.$$
Consider now a proper subgroup $\widehat{P}$ containing $\widehat{B}$. That is $\widehat{B}\leq \widehat{P}$. We call $\widehat{P}$ a {\it parabolic subgroup}. In analogy with the Langlands decomposition we have
$$\widehat{P}=M\widehat{A}_P\widehat{N}_P,$$
where $\widehat{N}_P\leq \widehat{N}$ and $M$ is finite dimensional and semisimple. Let $\widehat{A}=\widehat{A}_B$, then $\widehat{A}$ is finite dimensional with positive real diagonal entries and we have Iwasawa decomposition 
$$\widehat{G}=\widehat{K}\widehat{A}\widehat{N}$$
with uniqueness of expression.

In general
$$\widehat{G}=\widehat{K}M\widehat{A}_P\widehat{N}_P.$$
Then elements of $\widehat{G}$ may not have a unique expression, but the $\widehat{A}_P$-component, say $a\in \widehat{A}_P$, is uniquely determined ([Ga2], [Ga3]). 

Set
$$K_M=\widehat{K}\cap M$$
$$\Gamma_M=\widehat{\Gamma}\cap M.$$ 
Since $M$ is finite dimensional, we may choose  a cusp form $\phi$ on $K_M\backslash M/\Gamma_M$. Let $\Phi_s$ be a quasi--character on $\widehat{A}_P$, that is, a continuous map to the multiplicative group $\C^{\times}$
$$\Phi_s:\widehat{A}_P\to \C^{\times}.$$
We now extend $s$ to $\widehat{G}$ in the following way. Set
$$\Phi_{s}(g)=a^{s}\phi(m)$$
for $m\in M$, where $\phi$ is a cusp form on $K_M\backslash M/\Gamma_M$. We can now define Eisenstein series relative to a cusp form $\phi$ ([GMP]). Set
$$E_{\phi,s}(g)=\sum_{\gamma\in \widehat{\Gamma}/\widehat{\Gamma}\cap \widehat{P}} \Phi_{s}(ge^{-rD}\gamma)$$
where $r\in \R_{>0}$ and $D$ is the degree operator.

\medskip \noindent Then for $Re(\Phi_s(h_i))<-2$,  $E_{\phi,s}(g)$ converges uniformly and absolutely on compact sets, where $h_i$ are the simple coroots. As we comment in Section~\ref{MinMax}, convergence for Eisenstein series over the minimal parabolic subgroup $B$ implies convergence for Eisenstein series over $P$.

\subsection{Eisenstein series on non--affine Kac--Moody groups}\label{KMEseries}

 Let $G$ be a non-affine Kac--Moody group associated to a Kac--Moody algebra $\mathfrak{g}$ with symmetric generalized Cartan matrix. Let $g=kan\in G(\R)$, written in Iwasawa form. Let $s\in {\mathbb C}^{\times}$. We let $a\in A$ and define
$$\Psi_{s}:A\longrightarrow {\mathbb C}^{\times}$$
$$a\mapsto a^s.$$  
Extend $\Psi_{s}$ to $G$:
$$\Psi_{s}:G\longrightarrow {\mathbb C}^{\times}$$
$$\Psi_{s}(kan)\mapsto \Psi_{s}(a),$$
which is well defined since the Iwasawa decomposition is unique. Then $\Psi_{s}$ is left $K$--invariant and right $N$--invariant.  Let $\Gamma=G(\Z)$.
Clearly $\Psi_{s}(g\cdot \gamma)=\Psi_{s}(g)$, $g\in G$, $\gamma\in\Gamma\cap N$, so $\Psi_s$ is constant on $\Gamma$--orbits.  

 When $G$ is of non--affine type, we consider Eisenstein series defined in this way, without reference to a cusp form. There are currently no known examples of cusp forms on non--affine Kac--Moody groups.

 Let $B$ denote the minimal parabolic subgroup of  $G(\R)$. Relative to a coherently ordered basis $\Psi$ for $V^{\lambda}_{\Z}$, $\Gamma$ has a representation in terms of infinite matrices with integral entries.

 For $g\in G$ define Eisenstein series on $G(\R)$ by
$$E_{s}(g)\quad:=\quad \sum_{\gamma\in( \Gamma/\Gamma\cap{B})}\quad 
\Psi_{s}(g\gamma).$$
This is defined in analogy with Eisenstein series on the rank 2 hyperbolic Kac--Moody groups as in [CGGL] and  [CLL] and is the analog of Eisenstein series on $SL_2(\R)$ as in Section~\ref{Eseries}.

  Now let $\mathfrak{h}$ be the Lie algebra of $A$ and let $h_i$ be the simple coroots.
We let $h\in \mathfrak{h}$ and define
$$\Psi_{s}(\mathfrak{h}):\mathfrak{h}\longrightarrow {\mathbb C}^{\times}$$
$$h\mapsto h^s.$$

In [CCCM], the authors show that the Eisenstein series  $E_{s}(g)$ converges almost everywhere inside a cone in the region $Re(\Psi_{s}((h_i))< -2$ for each $i$. We will not comment on the difficult question of meromorphic continuation to the whole complex plane here.

 Our definition applies to Kac--Moody groups such as $E_{10}$ and $E_{11}$. However, it is an open question to give a physical interpretation to these Eisenstein series.

\section{Properties of Eisenstein series}

\subsection{Minimal parabolic versus maximal parabolic Eisenstein series}\label{MinMax} Let $G$ be a finite dimensional simple algebraic group. Let $B$ be its Borel subgroup. Let $P$ be a parabolic subgroup containing ${B}$ and define Eisenstein series
$$E_{P,s}(g)\quad=\quad \sum_{\gamma\in( \Gamma/\Gamma\cap{P})}\quad 
\Psi_{s}(g\gamma)$$
relative to $P$. If $P$ is a maximal parabolic subgroup $P=P_i$ associated to the simple root $\alpha_i$, then in [GMRV], the authors show that the character stabilized by $P_i$ is $\Psi_{m\omega_i}$ where $\omega_i$ is the fundamental weight of node $i$ on the Dynkin diagram and $m\in\C$.

 The minimal parabolic Eisenstein series can be written as
$$E_{B,s}(g)=\sum_{\gamma\in {\Gamma}/{\Gamma}\cap {B} } \Psi_{s}(g\gamma)
=\sum_{\gamma_2\in {\Gamma}/{\Gamma}\cap {P}}\ \ 
\sum_{\gamma_1\in  {\Gamma}\cap {P}/{\Gamma}\cap {B}}
\Psi_{s}(g\gamma_2\gamma_1).$$
Choosing $s=m\omega_i$ makes the double sum collapse to a single sum over $\gamma_2$, yielding a maximal parabolic Eisenstein series.

  Now consider Eisenstein series constructed using representation theory as in Section~\ref{EseriesRep}. It follows that choosing the module to be  a fundamental module  results in a maximal parabolic Eisenstein series. However, if one chooses an arbitrary module, in `most' cases the resulting Eisenstein series   will be a minimal parabolic Eisenstein series.


 Now let $G$ be an affine Kac--Moody group. In [G4], equation (3.2), the author used the method Borel--Bernstein ([Bo], see also  [GMRV]) to write the minimal parabolic E-series as a double sum which uses any other parabolic subgroup
$$\sum_{\gamma\in \widehat{\Gamma}/\widehat{\Gamma}\cap \widehat{B} } \Psi_{s}(ge^{-rD}\gamma)
=\sum_{\gamma_2\in \widehat{\Gamma}/\widehat{\Gamma}\cap \widehat{P}}\ 
\sum_{\gamma_1\in  \widehat{\Gamma}\cap \widehat{P}/\widehat{\Gamma}\cap \widehat{B}}
\Psi_{s}(ge^{-rD}\gamma_2\gamma_1).$$
The convergence of the left hand side implies convergence of the right hand side and thus the sum $\sum_{\gamma_1\in  \widehat{\Gamma}\cap \widehat{P}/\widehat{\Gamma}\cap \widehat{B}} \Psi_{s}(ge^{-rD}\gamma_2\gamma_1)$ can be replaced by a function of $\gamma_2$. The right hand side can therefore be rewritten 
as  a single  sum over $\gamma_2\in\widehat{\Gamma}/\widehat{\Gamma}\cap \widehat{P}$.

 In this case and the general Kac--Moody case,  for  Eisenstein series constructed from representations  as in Subsection~\ref{fund}, choosing $s=m\omega_i$ corresponds to an Eisenstein series associated to the single fundamental module $V^{\omega_i}$. It is unclear in the general case if a particular choice of $s$ allows us to collapse the minimal parabolic Eisenstein series into a maximal parabolic Eisenstein series.


\subsection{The constant term}  When $G$ is finite dimensional, we define the {\it constant term} 
$$\int_{{N}/{\Gamma}\cap {N}} E_{s}(gu) du.$$
In this case, $N/N\cap\Gamma$ is compact. When $G=SL_2$, $\Gamma=SL_2(\Z)$, $N/N\cap\Gamma=S^1$. 
 The group $G$ has Bruhat decomposition $G = \sqcup_{w} G_w$ where
$$
G_w = BwB,
$$
$w\in W$ and $B$ is the `upper triangular', or Borel subgroup. Each Bruhat cell 
$$\Gamma / \Gamma \cap B = \sqcup_{w\in W} (\Gamma\cap G_w) / (\Gamma \cap B)$$ then contributes one term to the constant term. The number of terms in the constant term thus equals the cardinality of the Weyl group $W$.

 When $\widehat{G}$ is an affine Kac--Moody group, $W$ is infinite and the prounipotent radical $\widehat{N}$ of $\widehat{B}$ is infinite dimensional. However, the quotient $\widehat{N}/\widehat{\Gamma}\cap \widehat{N}$ is a projective limit  of finite dimensional nil manifolds and hence carries a measure which allows us to integrate over the quotient. In [G4], the author defined the constant term as 
$$\int_{\widehat{N}/\widehat{\Gamma}\cap \widehat{N}} E_{\phi,s}(ge^{-rD}u) du.$$

 In [FK], the authors showed that remarkably, for certain choices of the parameter $s$, there are only finitely many terms in the constant terms for $E_9$, $E_{10}$ and $E_{11}$. This uses  a reduction method of [GMRV] for eliminating terms of the constant term. For affine Kac--Moody algebras, this also uses the structure of the affine Weyl group and the Weyl group orbits.


  Now let the rank 2 group $G$ be constructed over a finite field $\mathbb{F}_q$. Then $G$ is a locally compact group with a well defined Haar measure. In [CGGL], the authors considered rank 2 hyperbolic Kac--Moody groups $G$ and  constructed Eisenstein series which converge absolutely in a half space and proved meromorphic continuation to the whole complex plane. They  obtained a {\it finite} Bruhat decomposition 
$$G\quad  =\quad  \mathcal{B}\sqcup \mathcal{B}w_{1}\mathcal{B}\quad  =\quad  \mathcal{B}\sqcup \mathcal{B}
w_{2}\mathcal{B}$$
where $\mathcal{B}$ is the stabilizer of the end of the fundamental apartment of the Tits building, which is a tree when $G$ has rank 2. This finite Bruhat decomposition is associated with a spherical building for $G$ with respect to a finite `spherical' Weyl group. This gives the  advantage of only having to integrate over 2 Bruhat cells, not infinitely many, in the computation of Eisenstein series and the constant term. Unfortunately such spherical $BN$--pairs, and hence such finite Bruhat decompositions,  do not exist for higher rank hyperbolic Kac--Moody groups.

In [CLL] the authors defined Eisenstein series on rank $2$ hyperbolic Kac--Moody  groups over $\R$, induced from quasi--characters. They proved convergence of the constant term and hence the almost everywhere convergence of the Eisenstein series.


\section{Conclusion and further directions}\label{conclusions}

 We constructed the non--affine Kac--Moody groups $G(\R)$ and $G(\Z)$ and defined Eisenstein series $E_{s}(g)$ on $K(G) \backslash G(\R)/ G(\Z)$.   The Eisenstein series $E_{s}(g)$ is invariant under translations in $G(\Z)$ and hence has a Fourier expansion. Determining the Fourier expansion and the constant term of the Eisenstein series is of interest both from mathematical and physical points of view.

 The Eisenstein series should be an eigenfunction of the Casimir operators including the Laplacian on $K(G) \backslash G(\R)/ G(\Z)$. It would be interesting to find the explicit forms of these eigenvalue equations using the higher order Casimir operators introduced in [K1]. So far, only the second order Casimir operator has been analyzed in detail.

 The coset space $K(G) \backslash G(\mathbb{R})$ is a symmetric space, on which one can define a differential operator $d$. The square of $d$ is then the Laplace operator $\Delta = d^2$. For finite dimensional Lie groups, Eisenstein series are eigenfunctions of this Laplace operator with the eigenvalue equation from [GMRV]
$$
\Delta^{K \backslash G} E^G(\lambda,g) = 2(<\lambda,\lambda>-<\rho,\rho>)E^G(\lambda,g),
$$
where $\rho$ is the Weyl vector. Via duality symmetry in supergravities, we obtain Laplace eigenvalue equations for the coefficients of the higher order derivative corrections. 
 
 On the other hand, the group $G$ is endowed with a second order Casimir operator
$$
C_2 = \sum_{\alpha \in \Delta_+} n_{\alpha}X_{-\alpha} X_{\alpha} + \sum_{i=1}^{\ell} h_i^2 + 2h\cdot \rho,
$$
which acts on functions on the symmetric space $K(G) \backslash G(\R)$ as an operator that coincides with the Laplace operator.

 For affine Kac--Moody algebra the second Casimir operator contains an additional linear operator [K2]
$$
\eqnlab{Casimir}
C_2 = \sum_{\alpha \in \Delta_+} n_{\alpha}X_{-\alpha} X_{\alpha} + \sum_{i=1}^{\ell} h_i^2  + 2 h \cdot \rho + cD
$$
where $c$ is the central element and $D$ is the degree operator.

 An interesting future direction is to find the Laplace eigenvalue equation for Kac--Moody algebras, and to work out the precise relationship between the Laplace operator and the quadratic Casimir operator. Even in the affine case, where a full theory of affine Kac--Moody symmetric spaces has been worked out ([F]), the relationship is not clear. Investigation in this direction was touched upon in [FK].

 In some cases, constraints are required on Eisenstein series defined using representation theory in order that these automorphic forms are eigenfunctions of the Laplacian. These constraints are not required for the basic representation of $\mathfrak{sl}_n$, while a quadratic constraint is required for the vector representation of $\mathfrak{so}(d,d)$. An open question is how this might generalize to the Kac--Moody case. A more general question is to determine how $V_{\Z}=\mathcal{U}_{\Z}\cdot v_1$ breaks up into $G(\Z)$--orbits.

\newpage


\appendix

\section{Automorphic forms in supergravity theories}


Starting from eleven dimensional supergravity, the so--called {\it maximal supergravities} are obtained by toridal compactifications. After dimensionally reducing the classical action of eleven dimensional supergravity on an $n$--torus, the resulting $(11-n)$ dimensional supergravity action in the Einstein frame manifestly exhibits a coset symmetry of the form $K(E_{n(n)}(\R)) \backslash E_{n(n)}$.  In particular, the scalar fields of the maximal supergravity theory in $(11-n)$ dimensions take values in the coset $K(E_{n(n)}(\R)) \backslash E_{n(n)}$.

  These coset symmetries were first established for $1 \le n \le 8$ in [CJ]. In those cases, the symmetry groups $E_{n}(\R)$ are the split real forms of the simple exceptional Lie groups. The local symmetry groups $K(E_n)$ are the subgroups invariant under the Cartan involution. 

In dimensionally reduced supergravity in two dimensions, a study of its equations of motion revealed affine coset symmetry  $K(E_9) \backslash E_{9}(\R)$ ([Ju1] and [N]). 

Later, the hyperbolic Kac--Moody group $E_{10}$ was conjectured to be a symmetry of eleven dimensional supergravity ([DHN1]). The Lorentzian Kac--Moody group $E_{11}$ was also proposed to govern the dynamics of the full eleven dimensional supergravity ([W1]).

 The full list of symmetry groups in maximal supergravity theories is summarized in Table \ref{tab:coset}.
\begin{table}
  \begin{center}
    \begin{tabular}{l|lll}
      Dimension & $K(G)$              					& $G(\R)$                     & $G(\Z)$ \\
      \hline
      10, IIA   & \textbf{1}          					& $\R^+$          				& \textbf{1} \\
      10, IIB   & $SO(2)$             					& $SL_2(\R)$                  & $SL_2(\Z)$ \\
      9         & $SO(2)$      		  					& $SL_2(\R)\times \R^+$ 		& $SL_2(\Z)$ \\
      8         & $SO(3)\times SO(2)$ 					& $SL_3(\R)\times SL_2(\R)$ 	& $SL_3(\Z)\times SL_2(\Z)$ \\
      7         & $SO(5)$             					& $SL_5(\R)$                  & $SL_5(\Z)$ \\
      6         & $(Spin(5)\times Spin(5))/\Z_2$ 	& $Spin(5,5;\mathbb{R})$      & $Spin(5,5;\mathbb{Z})$ \\
      5         & $USp(8)/\Z_2$       					& $E_{6}(\mathbb{R})$         & $E_{6}(\mathbb{Z})$ \\
      4         & $SU(8)/\Z_2$        					& $E_{7}(\mathbb{R})$         & $E_{7}(\mathbb{Z})$ \\
      3         & $Spin(16)/\Z_2$     					& $E_{8}(\mathbb{R})$         & $E_{8}(\mathbb{Z})$ \\
		2         & $K(E_9)$            					& $E_{9}(\mathbb{R})$         & $E_{9}(\mathbb{Z})$ \\
		1         & $K(E_{10})$         					& $E_{10}(\mathbb{R})$        & $E_{10}(\mathbb{Z})$ \\
		0         & $K(E_{11})$         					& $E_{11}(\mathbb{R})$        & $E_{11}(\mathbb{Z})$
    \end{tabular}
  \end{center}
  \caption{Coset symmetries in maximal supergravities.}
  \label{tab:coset}
\end{table}

The first quantum corrections to the classical supergravity action were obtained by computing the scattering amplitudes. For instance, the four--graviton amplitude provides quantum corrections to the supergravity action of the form
$$
\ell_d^{8+2m-d} \int d^d x \sqrt{-G^{(d)}} \mathcal{E}_{(p,q)}^{(d)} \partial^{2m} \mathcal{R}^4,
$$
where $\ell_d$ is the $d$ dimensional Planck length and $G^{(d)}$ denotes the spacetime metric. The precise quartic tensor structure involving the Riemann curvature $\mathcal{R}$ has been suppressed in the above expression. More details of the scattering amplitudes are nicely summarized in [GMV].

It was then understood that in addition to the terms obtained by string perturbation theory, there also exist non--perturbative contributions named instanton corrections. In general, these are non--trivial to compute directly. The prime example is again the four--graviton scattering amplitude in maximal supergravity theories. In particular, the pure $\mathcal{R}^4$ correction term in type IIB supergravity is of the form [GG]
$$
\mathcal{E}_{(0,0)}^{(10)}(\tau,\bar{\tau}) = 2\zeta(3) \tau_2^{3/2} + 4\zeta(2) (\tau_2)^{-1/2} + 4\pi \sqrt{(\tau_2)} \sum_{N \neq 0} \mu_{-2}(N) N K_1(2\pi |N| (\tau_2)) e^{2\pi i N \tau_1},
$$
where the argument $\tau=\tau_1 + i\tau_2$ parameterizes the complex upper half plane $\mathbb{H}$. The first two terms constitute the constant term of the Eisenstein series $\mathcal{E}_{(0,0)}^{(10)}$. They correspond to the tree level and the one loop perturbative corrections respectively. The last part is an infinite sum of Bessel functions $K_1$ and accounts for the contributions from the D$(-1)$--instantons.

 The function $\mathcal{E}_{(0,0)}^{(10)}(\tau,\bar{\tau})$ is precisely the Eisenstein series for $SL_2$, written in its Fourier expanded form. In other words, the quantum corrections break the continuous coset symmetries of the maximal supergravity theories to the double coset $K(E_n) \backslash E_{n}(\R)/E_{n}(\Z)$. The corrections are encoded by the automorphic forms defined on $E_{n}(\Z)$.

 Later, it was discovered that the automorphic forms only account for the analytic part of the quantum corrections. There is an additional non--analytic part. Schematically, the scattering amplitudes of a maximal supergravity can be written as ([GMRV])
$$
A = A^{\text{analytic}} + A^{\text{non-analytic}}.
$$
The analytic part of the scattering amplitude in $d=(11-n)$ dimensions appears to be an expansion in automorphic forms $\mathcal{E}_{(p,q)}^{(d)}$ on the double coset $K(E_n) \backslash E_{n}(\R)/E_{n}(\Z)$. Each automorphic form is expanded in a Fourier expansion, which naturally separates the perturbative and non--perturbative quantum corrections. The constant term of the automorphic form can be computed using string perturbation theory. The remaining parts of the Fourier expansion of the automorphic forms have an interpretation as non--perturbative instanton corrections. For a generic correction term in $d$ dimensions, the corresponding coefficient is argued to be a transforming automorphic form.

 The full list of the so called U--duality groups $E_{n}(\Z)$ was given in [HT], where proof was provided for the cases $1 \le n \le 7$. In the same article the Kac--Moody integral forms $E_9(\Z)$ and $E_{10}(\Z)$ were also conjectured to be the U--duality groups in two and one dimensions, respectively. 

 Automorphic forms on $SL_n(\Z)$ and $Spin(d,d)$  in the context of supergravity theories were investigated in [OP2]. In particular,  the relation with the BPS states was discussed.

 In [GRV], using duality arguments, the coefficients in front of the $\mathcal{R}^4$, $\partial^4 \mathcal{R}^4$ and $\partial^6 \mathcal{R}^4$ corrections were identified as the functions $\mathcal{E}_{(0,0)}^{(d)}$, $\mathcal{E}_{(1,0)}^{(d)}$ and $\mathcal{E}_{(0,1)}^{(d)}$, respectively. The Laplace eigenvalue equations of these automorphic functions were found. The Fourier expansion of $\mathcal{E}_{(p,q)}^{(d)}$ has been analyzed in a series of articles starting from [GMRV].

  For $E_1=A_1$, $E_4$, $E_5$, $E_6$, $E_7$, $E_8$,
$$\mathcal{E}_{(0,0)}^{(d)}=2\zeta(3)E_{s}^{E_n}$$
where $s=3\omega_1=2\frac{3}{2}\omega_1$. For $E_1=A_1$, $E_6$, $E_7$, $E_8$,
$$\mathcal{E}_{(1,0)}^{(d)}=\zeta(5)E_{s}^{E_n}$$
where $s=5\omega_1=2\frac{5}{2}\omega_1$. The labeling of the nodes in the Dynkin diagram can be seen in Figure \ref{fig:DynkinEn}. The functions $\mathcal{E}_{(0,0)}^{(d)}$ and $\mathcal{E}_{(1,0)}^{(d)}$ in other cases involve linear combinations of Eisenstein series ([GMRV]). The constituent Eisenstein series in each of these combinations obey the same Laplace eigenvalue equation, due to the fact that their infinitesimal characters coincide up to a Weyl group transformation ([P]). The function $\mathcal{E}_{(0,1)}^{(d)}$ is more complicated and satisfies an inhomogeneous Laplace equation.
 
 Instead of computing the scattering amplitudes in lower dimensions directly, [LW] and its follow up works consider dimensional reductions of generic correction terms in ten dimensional supergravity theories. The $K(G) \backslash G(\R)$ coset formulation of the reduced theory is an important ingredient of this analysis. For simplicity, only the diagonal part of the coset representative is taken into account, while ignoring the exact tensor structure of the correction terms in the lower dimensional supergravities. However, this analysis still manages to put interesting constraints on the automorphic forms that potentially appear in supergravities.

 The authors of [GLW] and [GW1] suggest thus that at least one of the Eisenstein series that occur in supergravity theories should be constructed from the representation of $E_n$ with the highest weight $\omega_1$. A similar approach was also used in [BCN] to investigate properties of automorphic forms in $\mathcal{R}^2$, $\mathcal{R}^3$ and $\mathcal{R}^4$ corrections.

 The authors of [GW2] suggested that symmetries of all the maximal supergravity theories can be obtained from the Dynkin diagram of $E_{11}$ by removing one of its nodes (see also [RW], [BDN]). More specifically, maximal supergravity in $(11-n)$ dimensions has $E_n$ symmetry due to the internal torus and $A_{10-n}$ spacetime symmetry.

 The study of automorphic forms in supergravities has focused on automorphic functions so far, especially those appearing in front of $\partial^{2m} \mathcal{R}^4$ corrections. By taking into account supersymmetry, it has been shown that transforming automorphic forms appear in supergravities. In general, these automorphic forms transform as representations of the local symmetry group $K(G)$. 

 The relation between supersymmetry and transforming modular forms has been well studied, for example in [GGK] and [GS]. The transforming automorphic forms have been shown to be required from the analysis of dimensionally reducing higher order derivative corrections ([BCN]). However, the mathematical theory behind the transforming automorphic forms on  integer forms of higher rank groups and Kac--Moody groups is considerably more involved.

Although there are indications that automorphic forms based on Kac--Moody groups may play an important role in M--theory, their precise role is not yet understood. From the physical point of view, there are certain obstacles to overcome. One such obstacle appears in the study of $E_9(\Z)$ in 2 dimensions. Here, the supergravity action  cannot be cast in the Einstein frame, that is, the coordinate frame where the coset symmetries are manifest. It would be desirable to interpret Eisenstein series as coefficients of higher derivative corrections in the string scattering amplitudes. However, without the supergravity action in the Einstein frame, identifying coset symmetries in the string scattering amplitudes becomes more complicated.





\bigskip

\bigskip

\smallskip\noindent{Ling Bao}

\smallskip\noindent{Fundamental Physics, Chalmers University of Technology}

\smallskip\noindent{SE--412 96,  G\"oteborg, Sweden}

\smallskip\noindent{\it lingbao.work@gmail.com}
\bigskip

\smallskip\noindent{Lisa Carbone}

\smallskip\noindent{Department  of Mathematics, Rutgers University}

\smallskip\noindent{110 Frelinghuysen Rd, Piscataway, NJ 05584,  USA}

\smallskip\noindent{\it lisa.carbone@rutgers.edu}

\end{document}